\documentclass[twocolumn,showpacs,preprintnumbers,amsmath,amssymb,floatfix]{revtex4-1}
\usepackage{graphicx}
\usepackage{afterpage}
\usepackage{amsmath}
\usepackage{dcolumn}
\usepackage{bm}

\begin{document}




\title{Cascade coherence transfer and magneto-optical resonances at 455 nm excitation of cesium}




\author{M.~Auzinsh}%
\email{Marcis.Auzins@lu.lv}
\author{R.~Ferber}%
\author{F.~Gahbauer}%
\author{A.~Jarmola}%
\author{L.~Kalvans}%
\affiliation{The University of Latvia, Laser Centre, Rainis Blvd., LV-1586 Riga, Latvia}%
\author{A.~Atvars}%
\affiliation{Institute of Physical Research and Biomechanics, Maskavas Str. 22-1, LV-4604 Rezekne, Latvia}

\date{\today}

\begin{abstract}
We present an experimental and theoretical study of nonlinear magneto-optical resonances observed in 
the fluorescence to the ground state from the $7P_{3/2}$ state of cesium, which was populated directly by laser radiation 
at 455 nm, and from the $6P_{1/2}$ and $6P_{3/2}$ states, 
which were populated via cascade transitions that started from the $7P_{3/2}$ state and passed through various 
intermediate states. The laser-induced fluorescence (LIF) was observed as the magnetic field was scanned through zero.
Signals were recorded for the two orthogonal, linearly polarized components of the LIF. We 
compared the measured signals with the results of calculations from a model that was based on the optical Bloch 
equations  and averaged over the Doppler profile. 
This model was adapted from a model that had been developed for $D_1$ and $D_2$ excitation of alkali metal atoms. 
The calculations agree quite well with the measurements, 
especially when taking into account the fact that some experimental parameters were only estimated in the model. 
\end{abstract}





\maketitle

\section{\label{Intro:level1}Introduction}
The technique of populating atomic states via cascade transitions from higher-lying states has been used for many 
years to study atomic properties and quantum phenomena. The excited-state Hanle effect, or zero-field level crossing 
in the case of weak excitation, has been used to measure lifetimes and hyperfine structure (hfs) parameters of atomic 
states. For example, Tsukakoshi and Shimoda~\cite{Tsukakoshi:1969} and 
Carrington~\cite{Carrington:1972} used discharge lamps to observe the cascade Hanle effect in order to study decay times 
of atomic levels in xenon and neon, respectively. 
The cascade Hanle effect has been used to study lifetimes of alkali metal $S$ states~\cite{Bulos:1976} and, together with 
other cascade techniques, to measure lifetimes and hfs parameters of atomic states in alkali metal $D$ states~\cite{Tai:1975}. 
In some cases, the motivation for populating states via cascade transitions was
to populate states that were otherwise unreachable via direct excitation~\cite{Chang:1971,Gupta:1972b}.
Meanwhile, the ground-state Hanle effect was first observed 
by Lehmann and Cohen-Tannoudji~\cite{Lehmann:1964}. Schmieder~\cite{Schmieder:1970} and later 
Alzetta~\cite{Alzetta:1976} observed dark resonances, where the fluorescence is at a minimum at zero magnetic field, 
when exciting $D_1$ or $D_2$ transitions in alkali metal atoms by means of discharge lamps. 
Similar resonances were observed by means
of laser excitation by Ducloy \textit{et al.}~\cite{Ducloy:1973} in fluorescence signals and Gawlik 
\textit{et al.}~\cite{Gawlik:1974} in connection with the nonlinear Faraday effect. Much later, Dancheva 
\textit{et al.}~\cite{Dancheva:2000} observed bright resonances, which have a fluorescence maximum at zero magnetic field, 
in the $D_1$ and $D_2$ transitions of rubidium atoms in a vapor cell. Recently,  
Gozzini and co-workers observed the narrow magneto-optical resonances associated with the 
ground-state Hanle effect in the fluorescence from states that were populated by cascade transitions 
from higher-lying states~\cite{Gozzini:2009}. 
They excited the second resonance line of potassium with linearly and circularly 
polarized light and observed nonlinear magneto-optical resonances in the unpolarized 
fluorescence from the $4P_{1/2}$ and $4P_{3/2}$ transitions, which had been populated from the $5P_{3/2}$ state 
via spontaneous cascade transitions through various intermediate states. 
Measurements were obtained at various temperatures, but no theoretical description was given.  

In the present article, we describe an experimental study of nonlinear magneto-optical resonances 
observed in the fluorescence to the ground-state via various de-excitation pathways from the $7P_{3/2}$ state 
(second resonance line) of cesium together with theoretical calculations to describe the observed signals. In addition, we
monitor the transfer of coherence through these cascades by measuring the polarization degree of the fluorescence radiation,
observed after excitation with linearly polarized radiation, and compare these measurements with theoretical calculations. 
Observations of nonlinear magneto-optical resonances in the fluorescence 
from states that are populated via cascades could be particularly interesting for 
magnetometry, because the resonances are narrow and can be observed at a wavelength far removed from the wavelength of 
the exciting laser radiation, which is the main source of noise in such measurements. Therefore, it seemed important to 
be able to study a system both experimentally and theoretically.   

The basic theory of the fluorescence from a state populated from above via cascade transitions to a state other than the 
ground state was given by Gupta et al.~\cite{Gupta:1972} for linear excitation. 
The theory was based on the optical Bloch equations for the density matrix. 
In 1978 Picqu\'e used the optical Bloch equations to describe accurately dark resonances that arose in nonlinear 
excitation of one hyperfine component of the $D_1$ transition in a strongly excited 
beam of sodium atoms~\cite{Picque:1978}. In recent years, such 
models have achieved very good agreement for the $D_1$ transitions of 
cesium~\cite{Auzinsh:2008} and rubidium~\cite{Auzinsh:2009} when averaging over the Doppler profile and taking into 
account the coherence properties of the laser radiation, as well as all adjacent hyperfine states and even the small 
effect of the mixing of magnetic sublevels in the magnetic field. 

\begin{figure}[htbp]
	\centering
		\resizebox{\columnwidth}{!}{\includegraphics{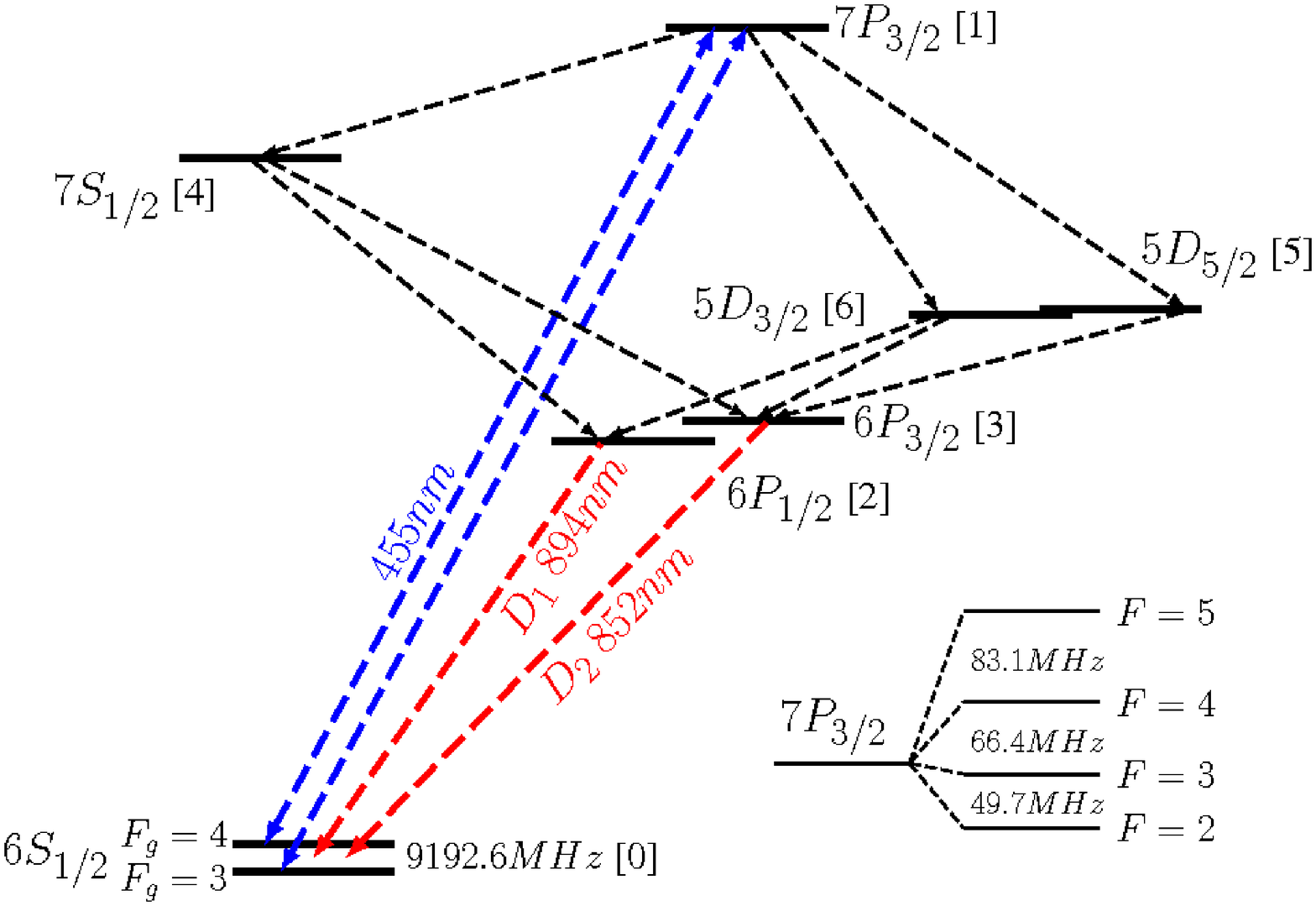}}
	\caption{\label{fig:levels} (Color online) Level diagram. Excitation takes place at 455 nm. Fluorescence is 
observed at 455 nm, 894 nm, and 852 nm. The numbers in square brackets correspond to the labeling scheme used in the 
equations of Sec.~\ref{section:rate}. 
}
\end{figure}

In the context of the present study it was necessary to adapt the theoretical model developed for
magneto-optical resonances in the $D$ lines of alkali metal atoms under nonlinear excitation 
in Refs.~\cite{Auzinsh:2008,Auzinsh:2009} to the cascade
transitions that result when the second resonance line of alkali atoms was excited. 
We compared experimentally measured signals with the results of calculations of the intensity of 
direct flourescence from the $7P_{3/2}$ state of cesium, as well as fluorescence from the 
$6P_{3/2}$ state ($D_2$ line) and the $6P_{1/2}$ state ($D_1$ line). The calculations were based on an extension of 
a theoretical model developed for the first resonance line of alkali metal atoms. However, in the case of cascade 
transitions, the large number of decay channels leads to a density matrix that is substantially larger than in the 
case for the first resonance line. 
We studied the unpolarized fluorescence intensity emitted 
along the direction of the scanned magnetic field as well as the 
polarized fluorescence intensity and the polarization degree. 
Figure~\ref{fig:levels} shows the atomic states involved in our experiment. The figure 
includes the exciting line, the cascade pathways, and the observed fluorescence lines. The theoretical model took
into account the population and coherence transfer of all possible de-excitation paths. As a result, 
it was necessary to solve very large systems of equations, which is computationally intensive and, thus, time-consuming. 
Therefore, instead of searching for the optimal parameters needed to describe the experimental signals in detail, 
we aimed to reproduce and understand the experimental features using estimated values for the model 
parameters.

\section{\label{Experiment:level1} Experimental Description}
In the experiment, a Toptica TA-SHG110 laser at 455 nm was used to excite cesium atoms in a vapor cell. The cell
was home-made and kept at room temperature at the center of a three-axis Helmholtz coil system. Two sets of coils
were used to cancel the ambient magnetic field, while the third set was used to scan the magnetic field $B$
from -7 Gauss to +7 Gauss by means of a Kepco BOP-50-8M bipolar power supply.  
The laser was usually tuned to the frequency for which the fluorescence at zero magnetic field  was at a maximum for 
a given transition, except in the case of certain studies where it was deliberately detuned from this frequency by a known
amount. The laser frequency was monitored using a High-Finesse WS7 wavemeter to
ensure that the frequency did not drift significantly during an experiment. During a given measurement, the laser 
frequency did not drift by more than 10 MHz.

\begin{figure}[htbp]
	\centering
		\resizebox{0.5\columnwidth}{!}{\includegraphics{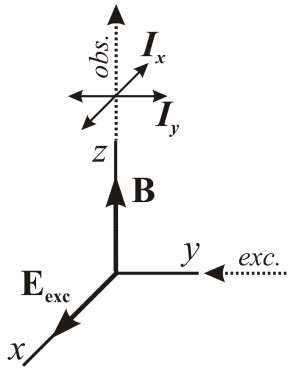}}
	\caption{\label{fig:geometry} Experimental geometry. The relative orientation of the laser 
beam (\textit{exc}), laser light polarization (\textbf{E$_{exc}$}), magnetic field (\textbf{B}), 
and observation direction (\textit{obs}) are shown. $I_x$ and $I_y$ are the linearly polarized components of 
the LIF intensity.  
}
\end{figure}

The geometry of the polarization vector of the exciting laser radiation, the magnetic field, and the direction
of fluorescence observation are given in Fig.~\ref{fig:geometry}. The fluorescence light was focused with a lens
onto a polarizing beam splitter, which directed two orthogonally polarized components of the fluorescence radiation
to two separate photodiodes (Thorlabs FDS-100). In front of the polarizing beam-splitter, interference filters were used
to select fluorescence at 455 nm, 852 nm, or 894 nm. Two different polarizing beam splitters were used, depending on the 
wavelength of the fluorescence radiation being observed: one was used for observations at 852 and 894 nm, while another
was used for observations at 455 nm. The photodiode signals were amplified and recorded separately on an Agilent DSO5014A
oscilloscope. To balance the amplifiers of the two photodiodes, the laser beam polarization was turned in such a way 
that the polarization vector of the laser radiation was parallel to the magnetic field. The difference signal ($I_x-I_y$) 
in this case should be zero when the amplifications of the photodiodes are properly balanced. 
Differences in sensitivity to unpolarized light and electronic offsets present in the absence of any light were also 
checked and taken into account. 

The cross-section of the laser beam determines the transit relaxation rate, and it was 3.2 mm$^2$. 
The beam cross-section was determined by considering the area of the beam 
where the power density was within 50\% of the maximum power density. The beam profile, which was approximately 
circular, was characterized by means of a Thorlabs BP104-VIS beam profiler.  
Different powers were selected by means of neutral density filters. 
Unless otherwise specified, the results presented in this article were obtained with a laser beam whose cross-sectional 
area was 3.2 mm$^2$ and whose total laser power was 40 mW. 
For some experiments, diminished laser powers were obtained using neutral density filters: 10 mW, 2.5 mW, and 0.625 mW. 
The signal background was determined by tuning the laser away from the resonance. 
No additional background from scattered laser-induced fluorescence (LIF) was taken into account in the analysis.

\section{\label{Theory:level1}Theoretical Model}
\subsection{Outline of the model}\label{section:outline}

In order to build a model of the nonlinear Hanle effect in alkali atoms
confined to a cell, we used the density matrix of an
atomic ensemble. The diagonal elements of the density matrix $\rho
_{ii}$ of an atomic ensemble describe the population of a certain
atomic level $i$, and the non-diagonal elements $\rho _{ij}$
describe coherences created between the levels $i$ and $j$. In our
particular case the levels in question are magnetic sublevels of a
certain hfs\ level. If atoms are excited from the ground state hfs
level $g$ to the excited state hfs level $e$, then the density
matrix consists of elements $\rho _{g_{i}g_{j}}$ and $\rho
_{e_{i}e_{j}}$, called Zeeman coherences, as well as
"cross-elements" $\rho _{g_{i}e_{j}}$, called optical coherences.

The time evolution of the density matrices is described by optical Bloch equations (OBEs), which can be written 
as~\cite{Stenholm:2005,Auzinsh:2010}:
\begin{equation}
i\hbar \dfrac{\partial \rho }{\partial t}=\left[ \widehat{H},\rho \right]
+i\hbar \widehat{R}\rho ,  \label{1}
\end{equation}%
where the operator $\widehat{R}$ represents the relaxation matrix.
If an atom interacts with laser light and an external \emph{dc}
magnetic field, the Hamiltonian can be expressed as $\widehat{H}=\widehat{H}_{0}+\widehat{H}_{B}+\widehat{%
V}$. $\widehat{H}_{0}$ is the unperturbed atomic Hamiltonian, which
depends on the internal atomic coordinates, $\widehat{H}_{B}$ is the
Hamiltonian of the atomic interaction with the magnetic field, and $\widehat{V%
}=-\widehat{\mathbf{d}}\cdot \mathbf{E}\left( t\right) $ is the interaction operator with the oscillating 
electric field in dipole approximation, where $\widehat{\mathbf{d}}$\ is the electric
dipole operator and $\mathbf{E}\left( t\right) $, the electric field
of the excitation light.

When using the OBEs to describe the interaction of alkali atoms with
laser radiation in the presence of a \emph{dc} magnetic field, we
describe the light
classically as a time dependent electric field of a definite polarization $%
\mathbf{e}$:%
\begin{equation}
\mathbf{E}\left( t\right) =\varepsilon \left( t\right) \mathbf{e}%
+\varepsilon ^{\ast }\left( t\right) \mathbf{e}^{\ast }  \label{2a}
\end{equation}%
\begin{equation}
\varepsilon (t)=\left\vert \varepsilon _{\overline{\omega }}\right\vert
e^{-i\Phi \left( t\right) -i\left( \overline{\omega }-\mathbf{k}_{\overline{%
\omega }}\mathbf{v}\right) t},  \label{2b}
\end{equation}%
where $\overline{\omega }$\ is the center frequency of the spectrum
and  $\Phi \left( t\right) $ is the fluctuating phase, which gives
the spectrum a finite bandwidth. In this model
the line shape of the exciting light is assumed to be Lorentzian with line-width
$\Delta \omega $. As each atom moves with a particular velocity $\mathbf{v}$, it experiences a shift $\overline{\omega }-%
\mathbf{k}_{\overline{\omega }}\mathbf{v}$ in the laser frequency due to the
Doppler effect, where $\mathbf{k}_{\overline{\omega }}$ is the wave vector
of the excitation light. The treatment of the Doppler effect is described in Sec. \ref{section:rate}.

The matrix elements of the dipole operator $\widehat{\bf{d}}$ that couple the $i$ sublevel with the $j$
sublevel can be written as: $d_{ij}=\langle i|\widehat{\mathbf{d}}\cdot
\mathbf{e}|j\rangle $. In the external magnetic field, sublevels are
mixed so that each sublevel $i$ with magnetic quantum number $M$ 
labeled as $\xi $ is a mixture of different
hyperfine levels $|F~M\rangle $ with mixing coefficients
$C_{i,F,M}$:
\begin{equation}
|i\rangle =|\xi M\rangle =\sum_{F}C_{i,F,M}|FM\rangle.   \label{3}
\end{equation}%
The mixing coefficients $C_{i,F,M}$ are obtained as the eigenvectors of the
Hamiltonian matrix of a fine structure state in the external magnetic field.
Since some of the upper states in the cascade system have rather small hyperfine splittings, 
the mixing of magnetic sublevels can be rather significant. For example, at a magnetic field of 7 Gauss, 
the mixing coefficients for the $7P_{3/2}$ state are on the order of 10\%, while in the $5D_{3/2}$ state 
the magnetic sublevels are fully mixed.

The dipole transition matrix elements $\langle
F_{k}M_{k}|\mathbf{d}\cdot \mathbf{e}|F_{l}M_{l}\rangle $ should be
expanded further using angular momentum algebra, including the
Wigner -- Eckart theorem and the fact that the dipole operator acts
only on the electronic part of the hyperfine state, which
consists of electronic and nuclear angular momentum (see, for example, 
Refs.~\cite{Auzinsh:2005,Auzinsh:2010}).

\subsection{Rate equations}\label{section:rate}

The rate equations for Zeeman coherences are developed by applying
the rotating wave approximation to the optical Bloch equations
with an adiabatic elimination procedure for the optical
coherences~\cite{Stenholm:2005} and then accounting realistically for the
fluctuating laser radiation by taking
statistical averages over the fluctuating light field phase (\emph{%
the decorrelation approximation}) and assuming a specific phase
fluctuation model: random phase jumps or continuous random phase
diffusion. As a result we arrive at the rate equations for Zeeman
coherences for the ground and excited state sublevels of atoms
\cite{Blush:2004}. In applying this approach to a case in which atoms are
excited only in the finite region corresponding to the laser beam
diameter, we have to take into account transit relaxation.

In Ref.~\cite{Blush:2004}, only resonant excitation at the $D$ lines was considered with one ground state and one excited state.
In the case of the cascade transitions considered here we have more than two 
states, and so they are denoted as follows (see Fig.~\ref{fig:levels}): the 6S$_{1/2}$ state is denoted by '[0]' and the part of the
density matrix related to this level is represented by $\rho^{[0]}$.  The 7P$_{3/2}$ state is denoted by '[1]' 
and the part of the density matrix that corresponds to it as $\rho^{[1]}$. Similarly, the 6P$_{1/2}$ state is indicated 
by '[2]', the 6P$_{3/2}$ state by '[3]', the 7S$_{1/2}$ state by '[4]', the  
5D$_{5/2}$ state by '[5]' and the 5D$_{3/2}$ state by '[6]'. 
If the above treatment of the OBEs is applied to the level scheme in discussion, 
we obtain the following rate equations:
\begin{widetext}
\begin{eqnarray}
\dfrac{\partial \rho^{[0]} _{g_{i}g_{j}}}{\partial t} &&=-i \omega
_{g_{i}g_{j}}\rho^{[0]} _{g_{i}g_{j}}-\gamma \rho^{[0]} _{g_{i}g_{j}}+\left(\underset{e_{k}e_{m}}{%
\sum }\Gamma _{[0] g_{i}g_{j}}^{[1] e_{k}e_{m}}\rho^{[1]} _{e_{k}e_{m}}+\underset{e_{k}e_{m}}{%
\sum }\Gamma _{[0] g_{i}g_{j}}^{[2] e_{k}e_{m}}\rho^{[2]} _{e_{k}e_{m}}+\underset{e_{k}e_{m}}{%
\sum }\Gamma _{[0] g_{i}g_{j}}^{[3] e_{k}e_{m}}\rho^{[3]} _{e_{k}e_{m}}\right) \notag \\
&&+\dfrac{\left\vert \varepsilon _{\overline{\omega }}\right\vert
^{2}}{\hbar ^{2}}\underset{e_{k},e_{m}}{\sum }\left(
\dfrac{1}{\Gamma _{R}+ i \Delta _{e_{m}g_{i}}}+\dfrac{1}{\Gamma
_{R}- i \Delta _{e_{k}g_{j}}}\right)
d_{g_{i}e_{k}}^{\ast }d_{e_{m}g_{j}}\rho^{[1]} _{e_{k}e_{m}}  \notag \\
&&-\dfrac{\left\vert \varepsilon _{\overline{\omega }}\right\vert ^{2}}{%
\hbar ^{2}}\underset{e_{k},g_{m}}{\sum }\left( \dfrac{1}{\Gamma
_{R}- i
\Delta _{e_{k}g_{j}}}d_{g_{i}e_{k}}^{\ast }d_{e_{k}g_{m}}\rho^{[0]} _{g_{m}g_{j}}+%
\dfrac{1}{\Gamma _{R}+ i \Delta _{e_{k}g_{i}}}d_{g_{m}e_{k}}^{\ast
}d_{e_{k}g_{j}}\rho^{[0]} _{g_{i}g_{m}}\right)  \notag \\
&&+\lambda\delta\left( g_{i},g_{j}\right)  \label{level0}
\end{eqnarray}

\begin{eqnarray}
\dfrac{\partial \rho^{[1]} _{e_{i}e_{j}}}{\partial t} &&=- i \omega
_{e_{i}e_{j}}\rho^{[1]} _{e_{i}e_{j}}- \left(\gamma + \Gamma^{[1]}\right) \rho^{[1]}_{e_{i}e_{j}} + \left(0\right)  \notag \\
&&+\dfrac{\left\vert \varepsilon _{\overline{\omega }}\right\vert
^{2}}{\hbar ^{2}}\underset{g_{k},g_{m}}{\sum }\left( \dfrac{1}{
\Gamma _{R}- i \Delta _{e_{i}g_{m}} }+\dfrac{1}{ \Gamma _{R}+ i
\Delta _{e_{j}g_{k}} }\right) d_{e_{i}g_{k}}d_{g_{m}e_{j}}^{\ast
}\rho^{[0]}
_{g_{k}g_{m}}  \notag \\
&&-\dfrac{\left\vert \varepsilon _{\overline{\omega }}\right\vert ^{2}}{%
\hbar ^{2}}\underset{g_{k},e_{m}}{\sum }\left( \dfrac{1}{\Gamma
_{R}+ i
\Delta _{e_{j}g_{k}}}d_{e_{i}g_{k}}d_{g_{k}e_{m}}^{\ast }\rho^{[1]} _{e_{m}e_{j}}+%
\dfrac{1}{\Gamma _{R}- i \Delta _{e_{i}g_{k}}}%
d_{e_{m}g_{k}}d_{g_{k}e_{j}}^{\ast }\rho^{[1]} _{e_{i}e_{m}}\right),
\label{level1}
\end{eqnarray}

\begin{eqnarray}
\dfrac{\partial \rho^{[2]} _{f_{i}f_{j}}}{\partial t} &&=- i \omega
_{f_{i}f_{j}}\rho^{[2]} _{f_{i}f_{j}}-\left(\gamma + \Gamma^{[2]}\right)
\rho^{[2]}
_{f_{i}f_{j}}\notag\\
&& + \left( \underset{e_{k}e_{m}}{%
\sum }\Gamma _{[2] f_{i}f_{j}}^{[4] e_{k}e_{m}}\rho^{[4]} _{e_{k}e_{m}} + \underset{e_{k}e_{m}}{%
\sum }\Gamma _{[2] f_{i}f_{j}}^{[6] e_{k}e_{m}}\rho^{[6]} _{e_{k}e_{m}}\right),
\label{level2}
\end{eqnarray}

\begin{eqnarray}
\dfrac{\partial \rho^{[3]} _{f_{i}f_{j}}}{\partial t} &&=- i \omega
_{f_{i}f_{j}}\rho^{[3]} _{f_{i}f_{j}}-\left(\gamma + \Gamma^{[3]}\right)
\rho^{[3]}
_{f_{i}f_{j}}\notag\\
&& +  \left(\underset{e_{k}e_{m}}{%
\sum }\Gamma _{[3] f_{i}f_{j}}^{[4] e_{k}e_{m}}\rho^{[4]} _{e_{k}e_{m}} +  \underset{e_{k}e_{m}}{%
\sum }\Gamma _{[3] f_{i}f_{j}}^{[5] e_{k}e_{m}}\rho^{[5]} _{e_{k}e_{m}}+ \underset{e_{k}e_{m}}{%
\sum }\Gamma _{f_{i}f_{j}}^{[6] e_{k}e_{m}}\rho^{[6]} _{e_{k}e_{m}}\right),
 \label{level3}
\end{eqnarray}

\begin{eqnarray}
\dfrac{\partial \rho^{[4]} _{f_{i}f_{j}}}{\partial t} &&=- i \omega
_{f_{i}f_{j}}\rho^{[4]} _{f_{i}f_{j}}-\left(\gamma + \Gamma^{[4]}\right)
\rho^{[4]}
_{f_{i}f_{j}} + \underset{e_{k}e_{m}}{%
\sum }\Gamma _{[4] f_{i}f_{j}}^{[1] e_{k}e_{m}}\rho^{[1]}_{e_{k} e_{m}} ,
\label{level4}
\end{eqnarray}

\begin{eqnarray}
\dfrac{\partial \rho^{[5]} _{f_{i}f_{j}}}{\partial t} &&=- i \omega
_{f_{i}f_{j}}\rho^{[5]} _{f_{i}f_{j}}-\left(\gamma + \Gamma^{[5]}\right)
\rho^{[5]}
_{f_{i}f_{j}} +  \underset{e_{k}e_{m}}{%
\sum }\Gamma _{[5] f_{i}f_{j}}^{[1] e_{k}e_{m}}\rho^{[1]}_{e_{k} e_{m}} ,
\label{level5}
\end{eqnarray}

\begin{eqnarray}
\dfrac{\partial \rho^{[6]} _{f_{i}f_{j}}}{\partial t} &&=- i \omega
_{f_{i}f_{j}}\rho^{[6]} _{f_{i}f_{j}}-\left(\gamma + \Gamma^{[6]}\right)
\rho^{[6]}
_{f_{i}f_{j}} + \underset{e_{k}e_{m}}{%
\sum }\Gamma _{[6] f_{i}f_{j}}^{[1] e_{k}e_{m}}\rho^{[1]}_{e_{k} e_{m}} .
\label{level6}
\end{eqnarray}
\end{widetext}

Here  $g_{i}$  denotes the ground state '0' magnetic sublevel, while
$e_{i}$ and $f_i$ denote magnetic sublevels of states '1', '2', '3', '4', '5', or '6' 
according to the associated index, with $e_i$ always referring to the level with higher energy. 
For example, $f_i$ in the expresion $\rho^{[5]} _{f_{i}f_{j}}$ belongs to level '5'. 
The term, $\Delta _{ij}=\bar{\omega}-\mathbf{k}_{\bar{\omega}}\mathbf{v}-\omega _{ij}$
expresses the actual laser shift away from the resonance for transitions between levels $\vert i \rangle$ 
and $\vert j \rangle$ for atoms moving with velocity \textbf{v}.
The total relaxation rate $\Gamma _{R}$ is given by 
$\Gamma _{R}=\frac{\Gamma^{[1]} }{2}+\frac{\Delta \omega }{2}+\gamma $, where
$\Gamma^{[k]} $ is the relaxation rate of the level 'k', $\gamma$ is
the  transit relaxation rate, and $\lambda$ is the
rate at which "fresh" atoms move into the interaction region. 
The rate $\gamma$ can be estimated as $%
1/(2\pi \tau )$, where $\tau$ is time it takes for an atom to cross the
laser beam at the mean thermal velocity $v_{th}$. It is assumed
that the atomic equilibrium density outside the interaction region
is normalized to $1$, which leads to $\lambda$ numerically equal to $\gamma$, since $\lambda=\gamma n_0$, 
where $n_0$ is the density of atoms. 
The term $\Gamma_{f_{i}f_{j}}^{e_{i}e_{j}}$ 
is the  rate at which excited state population and
coherences are transferred to the lower state as a result of
spontaneous transitions and it is obtained as follows~\cite{Auzinsh:2010}: 
\begin{widetext}
\begin{equation}
 \Gamma_{f_if_j}^{e_ie_j}=\Gamma^{s}(-1)^{2F_e-M_{e_i}-M_e}(2F_f+1) \sum_q\left(
\begin{array}{ccc}
 F_e     & 1 & F_f \\
-M_{e_i} & q & M_{f_j}
\end{array}
\right)
\left(
\begin{array}{ccc}
 F_e     & 1 & F_f \\
-M_{e_j} & q & M_{f_i}
\end{array}
\right)
\end{equation}
\end{widetext}

If the system is closed, all
excited state atoms return to the initial state through spontaneous transitions, $%
\underset{e_{i} f_{j}}{\sum }\Gamma _{[r] f_{j}f_{j}}^{[s] e_{i}e_{i}}= \alpha(s,r)\Gamma^{[s]} $.
where $\alpha(s,r)$ is the branching ratio of spontaneous emission from level 's' to level 'r'.
Furthermore, $\underset{r}{\sum }\alpha(s,r)=1$.

Equations (\ref{level0})---(\ref{level6}) describe the time evolution of the parts of the density matrix for 
states $[i] = [0]$---$[6]$, respectively. The first term on the right-hand side of each equation
describes the destruction of the Zeeman coherences due to magnetic sublevel splitting in an external magnetic 
field $\omega_{ij}=\left(E_i-E_j\right)/\hbar$. The
second term characterizes the effects of the transit relaxation rate ($\gamma^{[i]}$) and the spontaneous relaxation rate
($\Gamma^{[i]}$), with the latter being absent for the ground state '[0]'. The next term shows the transfer
of population and coherences from the upper state [$j$] to the state [$i$] described by a particular equation
due to spontaneous transitions; this term is equal to zero in equation (\ref{level1}), which describes  the '[1]' level,  
as no levels above this one are excited. For equations (\ref{level0}) and (\ref{level1}) the fourth term describes the 
population increase in the level due to laser-induced transitions, while the fifth term stands for the population driven 
away from the state via laser-induced transitions. Finally, the sixth term in equation (\ref{level0}) describes 
how the population of "fresh atoms" is supplied to the initial state from the volume outside the laser beam 
in a process of transit relaxation.

For a multilevel system that interacts with laser radiation, we can define the
effective Rabi frequency in the form $\Omega =\dfrac{\left\vert \varepsilon _{%
\overline{\omega }}\right\vert }{\hbar }\left\langle
J_{e}\right\Vert d\left\Vert J_{g}\right\rangle $, where $J_{e}$ is
the angular momentum of the excited state '1' fine structure level, and
$J_{g}$ is the angular momentum of the ground state '0' fine structure
level. The influence of the magnetic field appears directly in the
magnetic sublevel splitting $\omega _{ij}$ and indirectly in the
mixing coefficients $C_{i,F_{k},M_{i}}$ and $C_{j,F_{l},M_{j}}$ of
the dipole matrix elements $d_{ij}$.

We look at quasi-stationary excitation conditions so that $\partial
\rho^{[0]}_{g_{i}g_{j}}/\partial t=\partial \rho^{[1]} _{e_{i}e_{j}}/\partial t =\partial \rho^{[2]} _{f_{i}f_{j}}/\partial t = \partial \rho^{[3]} _{f_{i}f_{j}}/\partial t =\partial \rho^{[4]} _{f_{i}f_{j}}/\partial t =\partial \rho^{[5]} _{f_{i}f_{j}}/\partial t =\partial \rho^{[6]} _{f_{i}f_{j}}/\partial t =0$.

By solving the rate equations as an algebraic system of linear
equations for $\rho^{[0]} _{g_{i}g_{j}}$ and $\rho^{[1]} _{e_{i}e_{j}}$, $\rho^{[2]} _{e_{i}e_{j}}$,$\rho^{[3]} _{e_{i}e_{j}}$,$\rho^{[4]} _{e_{i}e_{j}}$,$\rho^{[5]} _{e_{i}e_{j}}$,$\rho^{[6]} _{e_{i}e_{j}}$  
we obtain the matrix of populations and Zeeman coherences 
for all levels involved ('0'---'6'). This matrix allows us to
obtain immediately the intensity of the observable fluorescence characterized
by the polarization vector $\mathbf{\tilde{e}}$~\cite{Auzinsh:2005,Auzinsh:2010}. 
Fluorescence that is transmitted from the excited-state level '$i$' to the ground-state level '$j$' is obtained as:
\begin{equation}
I^{[i]}(\mathbf{\tilde{e}})=\tilde{I_{0}}^{[i]}%
\sum_{g_{i},e_{i},e_{j}}d_{g_{i}e_{j}}^{(ob)\ast
}d_{e_{i}g_{i}}^{(ob)}\rho^{[i]} _{e_{i}e_{j}},  \label{intensity}
\end{equation}
where $\tilde{I_{0}}^{[i]}$ is a proportionality coefficient. The dipole
transition matrix elements $d_{e_{i}g_{j}}^{(ob)}$
characterize the dipole transition
from the excited state $e_{i}$ to some ground state $g_{j}$ for the
transition on which the fluorescence is observed.

To calculate the fluorescence 
produced by an ensemble of atoms, we have to treat the previously
written expression for the
fluorescence as a function of both the polarization vector of the fluorescence and the atomic velocity, 
$I^{[i]}(\mathbf{\tilde{e}})=I^{[i]}(\mathbf{\tilde{e}},\mathbf{k}_{\overline{\omega }}\mathbf{v})$ 
and average it over the Doppler profile while taking into account 
the different velocity groups $\mathbf{k}_{\overline{\omega }}\mathbf{v}$ with their respective statistical weights. 
If the unpolarized fluorescence without discrimination of the polarization or frequency
is recorded, one needs to sum the fluorescence over the
two orthogonal polarization components and all possible final state hfs levels.

\subsection{Model parameters}\label{section:parameters}

In order to perform theoretical simulations with the methods described in the previous section, a number of 
theoretical parameters and atomic constants had to be used. Some parameters are known rather precisely.
Thus, the hyperfine splitting constants for atomic levels involved 
in the cesium $D$ lines were obtained from Ref.~\cite{Steck:cesium}, while the magnetic dipole and electric 
quadrupole constants for the remaining energy levels were taken from Ref.~\cite{Arimondo:1977}. 
The natural linewidths for levels not involved in the $D$ lines were obtained from Ref.~\cite{Theodosiou:1984}. 
The branching ratios for the cascade transitions are available from the 
NIST database~\cite{Eriksson:1964,Kleiman:1962} and in Ref.~\cite{Kurucz:1995}.

For the parameters that were related to the experimental conditions, we used reasonable estimates 
based on measurements of the laser beam parameters and our previous experience.
The transit relaxation rate is the inverse of the mean time that 
an atom spends in the laser beam as it moves chaotically in the vapor cell with a thermal 
velocity. For a laser beam diameter of 2 mm (full width at half maximum of the intensity profile) 
and room temperature (293K), we used a value of 0.02 MHz. To estimate the 
Rabi frequency to be used in the simulations, we calculated the saturating laser power density 
for the excitation  transition using its natural linewidth and then related this value to the power densities used in the 
experiments. The saturating laser power density is the laser power density at which the stimulated emission equals 
the spontaneous decay rate, and it can be obtained from the formula~\cite{Alnis:2003}:
\begin{equation}
I_{sat}=\frac{4hc}{\lambda_{eg}^3} \frac{\Omega_{sat}^2}{\Gamma^{[1]}} 
\nonumber
\end{equation} 
In such a way the saturating Rabi frequency $\Omega_{sat}$ in our experiment was estimated to be about 5 MHz. 
In the calculations, the results were averaged over the Doppler profile with the appropriate weighting factor 
and a step-size of 2.5 MHz. Another parameter that had to be estimated was the laser frequency. For the experiment, the 
reference frequency for a given transition was the frequency at which maximum fluorescence was observed.

\section{\label{Results:level1}Results and Discussion}
Both orthogonal, linearly polarized polarization components $I_{x,y}$ were recorded in all experiments. 
To visualize the data, three quantities were considered: the unpolarized fluorescence ($I_x+I_y$), 
the polarized fluorescence ($I_x$ and $I_y$), and the polarization degree [$(I_x-I_y)/(I_x+I_y)$].  

\subsection{Unpolarized Fluorescence}\label{section:unpolarized}
\begin{figure*}[htbp]
	\centering
		\resizebox{8cm}{6cm}{\includegraphics{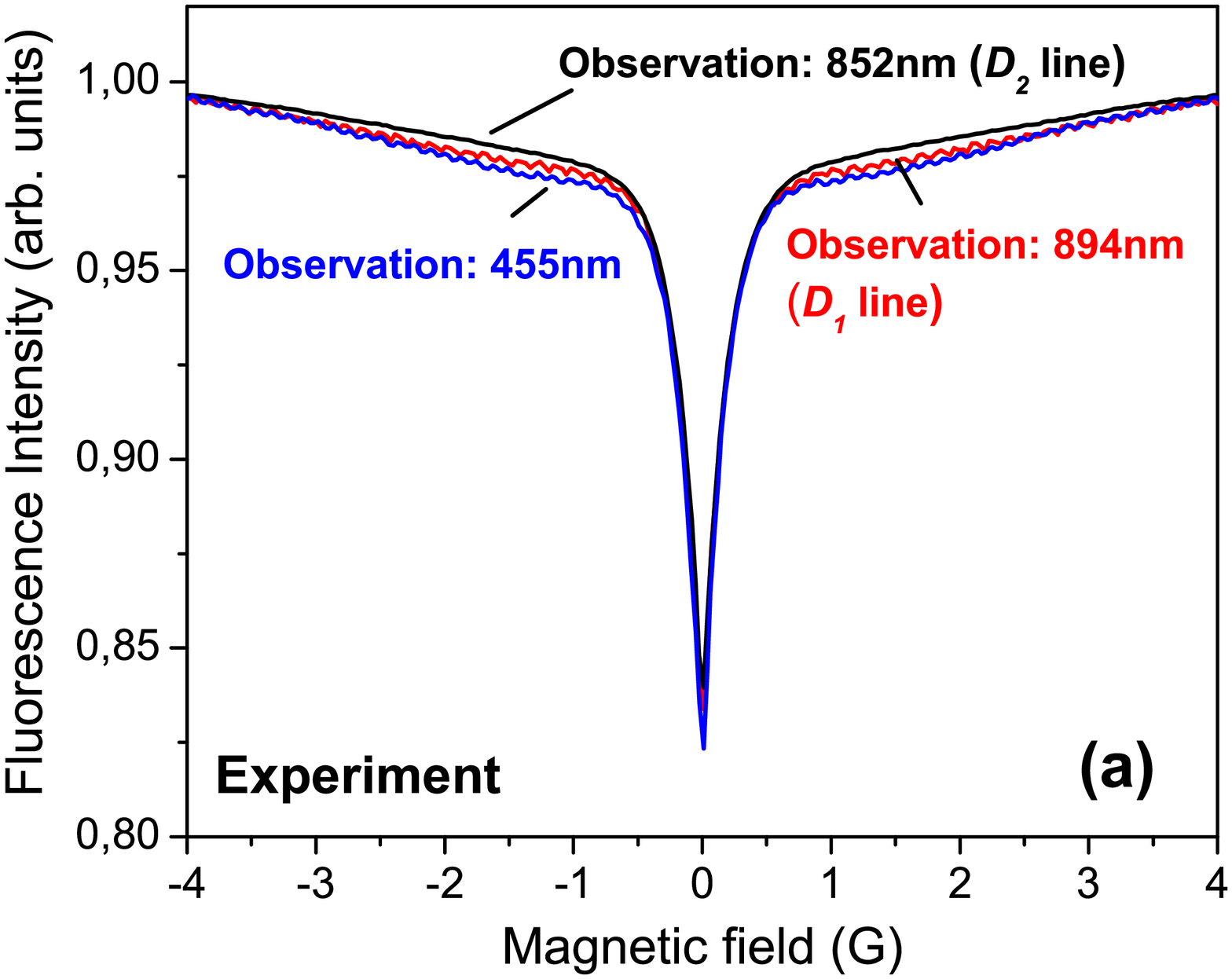}}
		\resizebox{8cm}{6cm}{\includegraphics{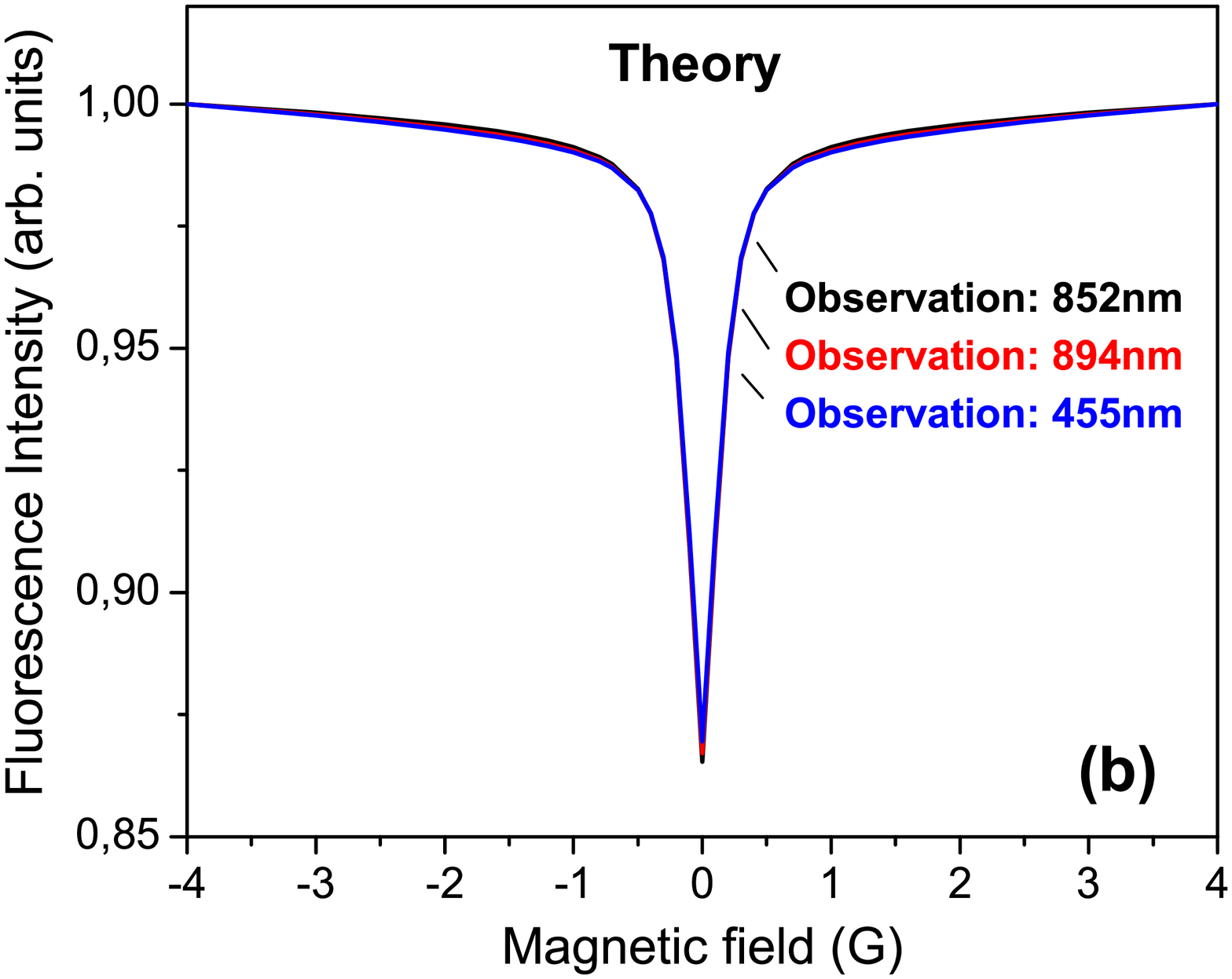}}
		\resizebox{8cm}{6cm}{\includegraphics{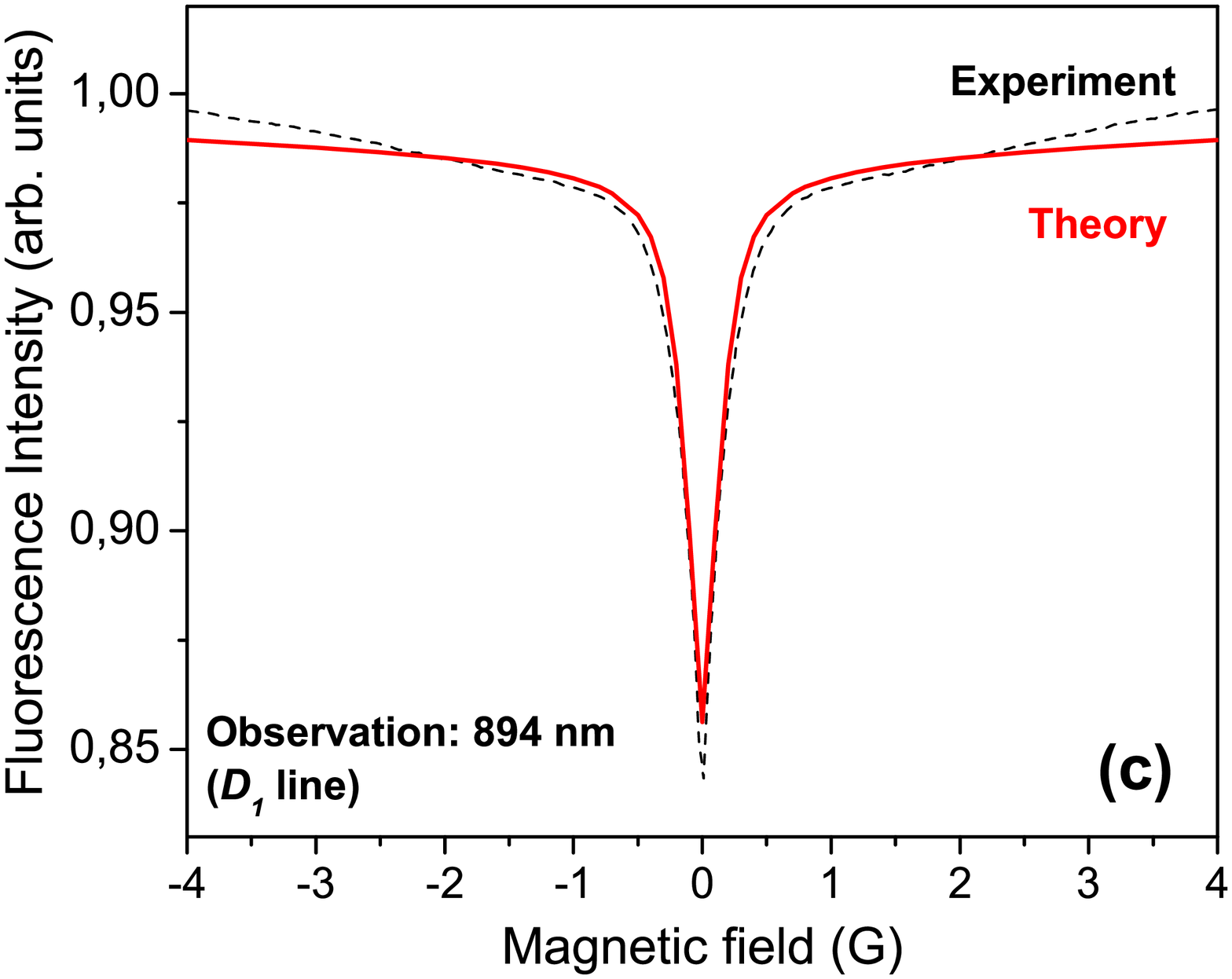}}
		\resizebox{8cm}{6cm}{\includegraphics{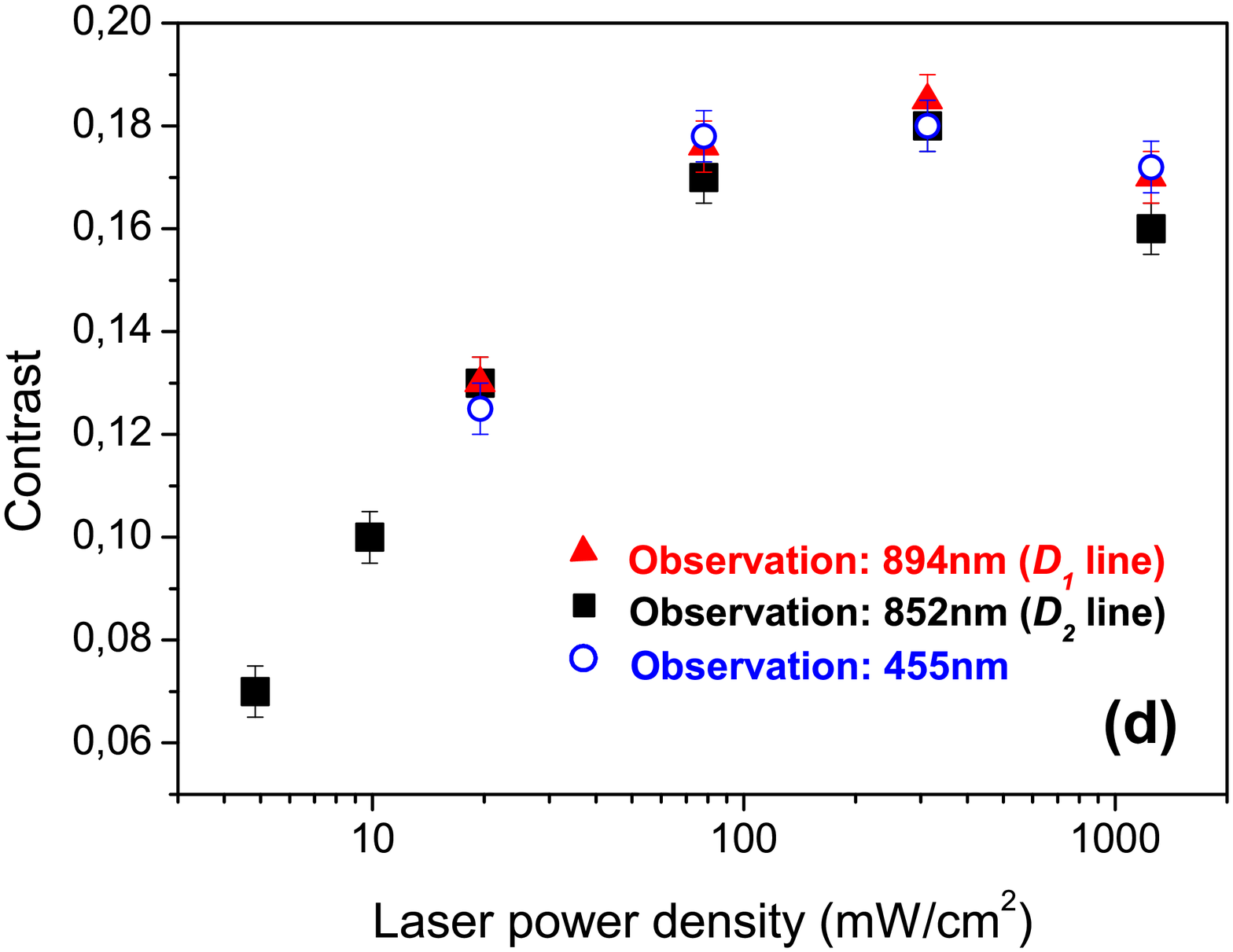}}
	\caption{\label{fig:fg3_np} (Color online) Intensity of the non-polarized fluorescence to the ground state 
versus magnetic field for excitation of the $6S_{1/2}(F_g=3)\rightarrow 7P_{3/2}$ transition at 455 nm. 
(a) Observed fluorescence from the $7P_{3/2}$ (direct), $6P_{3/2}$ (cascade), and $6P_{1/2}$ (cascade) states. 
The three curves are practically indistinguishable. 
(b) Theoretical calculations corresponding to the observations in (a). The three curves almost coincide.
(c) Experimental observation and theoretical calculation of the fluorescence from the $6P_{1/2}$ state ($D_1$ transition).
(d) Observed contrast as a function of the laser power density for the fluorescence from the three levels mentioned in (a).
}
\end{figure*}

Figure~\ref{fig:fg3_np} displays results obtained by exciting the $7P_{3/2}$ state from the $F_g=3$ 
ground-state level and observing direct, unpolarized fluorescence to the ground state from the $7P_{3/2}$ state as well as 
fluorescence to the ground state from the $6P_{3/2}$ ($D_2$ line) and $6P_{1/2}$ ($D_1$ line) states, which had been populated 
by cascades transitions. 
In all cases, a narrow, dark, Hanle-type resonance was observed. The experiment showed, and theoretical calculations 
confirmed, that the shape of the resonance does not depend on which fluorescence line to the ground state is observed
[see Fig.~\ref{fig:fg3_np}(a)] and Fig.~\ref{fig:fg3_np}(b)]. In fact, the three experimental curves in 
Fig.~\ref{fig:fg3_np}(a) are practically indistinguishable, and the same is true for the three theoretical curves in
Fig.~\ref{fig:fg3_np}(b). Fig.~\ref{fig:fg3_np}(c) shows the unpolarized fluorescence 
intensity observed from the $6P_{1/2}$ state populated via cascades versus magnetic field. 
Experimental measurements and results from calculations are shown on the same plot. Because the calculations are 
extremely time-consuming, it was not possible to vary the model parameters in order to find those experimental parameters 
that could not be measured directly. 
Nevertheless, by making reasonable estimates of the parameter values based on the experience gained in  
Refs.~\cite{Auzinsh:2008,Auzinsh:2009}, it was possible to obtain almost perfect agreement between experiment and 
theory (see Sec.~\ref{section:parameters}). 
The theoretical calculation assumed that the laser frequency was tuned to the $F_g=3\rightarrow F_e=3$ transition. 
 
Figure~\ref{fig:fg3_np}(d) shows the measured resonance contrasts for each observed fluorescence line as a function of 
laser power density. Similar to the case of nonlinear magneto-optical resonances in $D$ line excitation, 
the contrast increased with increasing laser power density up to some maximum and then decreased 
(see, for example, Fig.~8 in Ref.~\cite{Auzinsh:2009}).  

\begin{figure*}[htbp]
	\centering
		\resizebox{8cm}{6cm}{\includegraphics{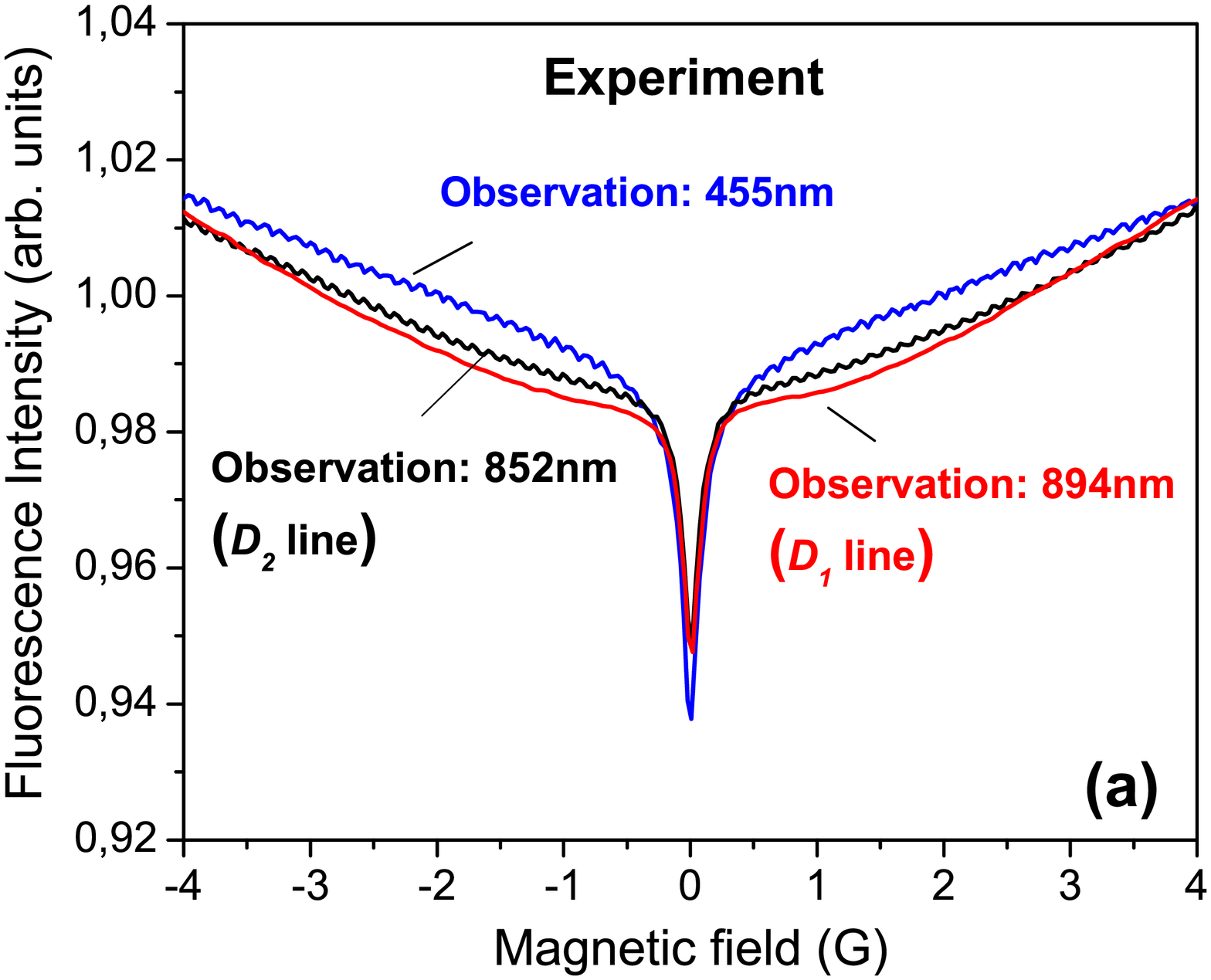}}
		\resizebox{8cm}{6cm}{\includegraphics{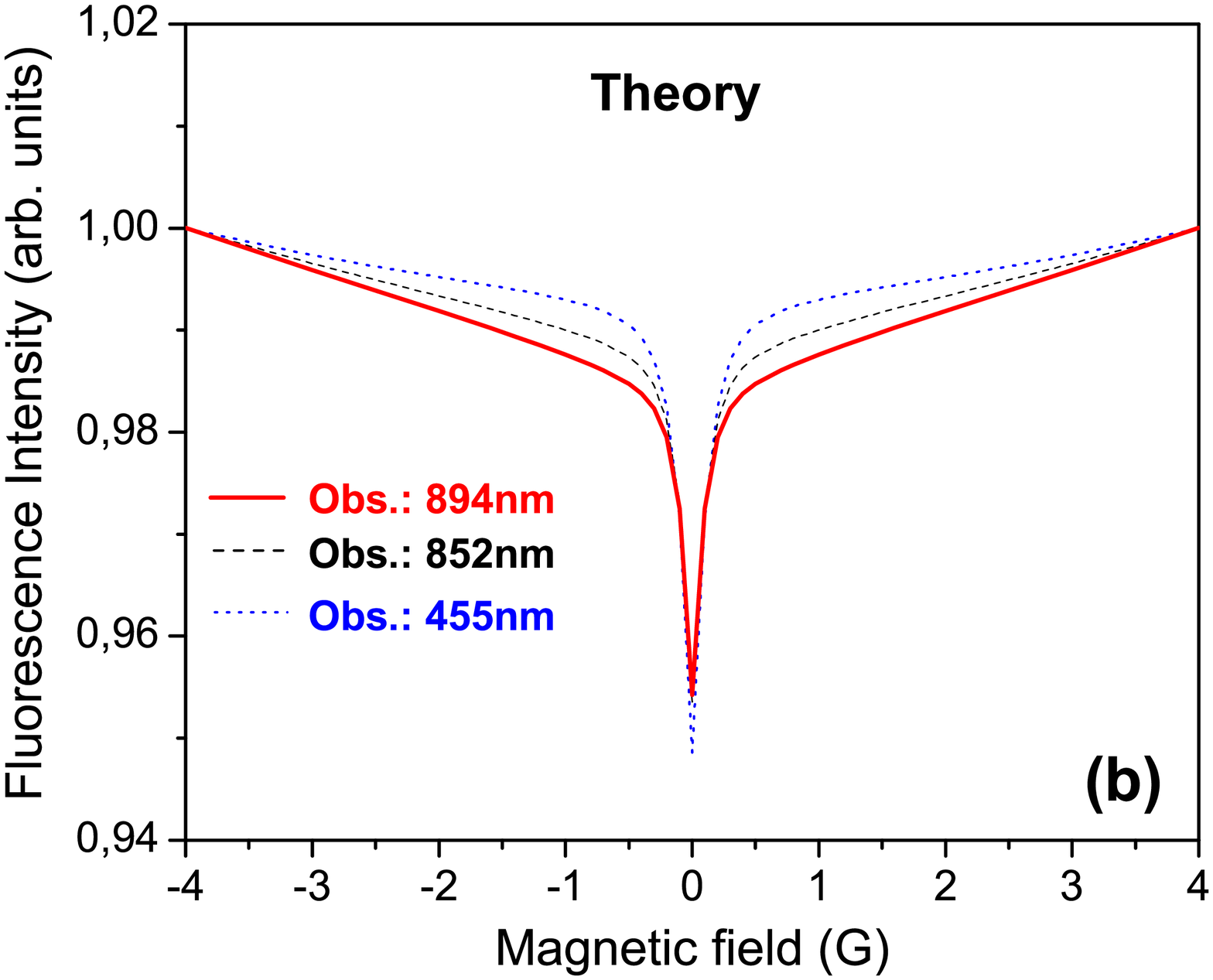}}
		\resizebox{8cm}{6cm}{\includegraphics{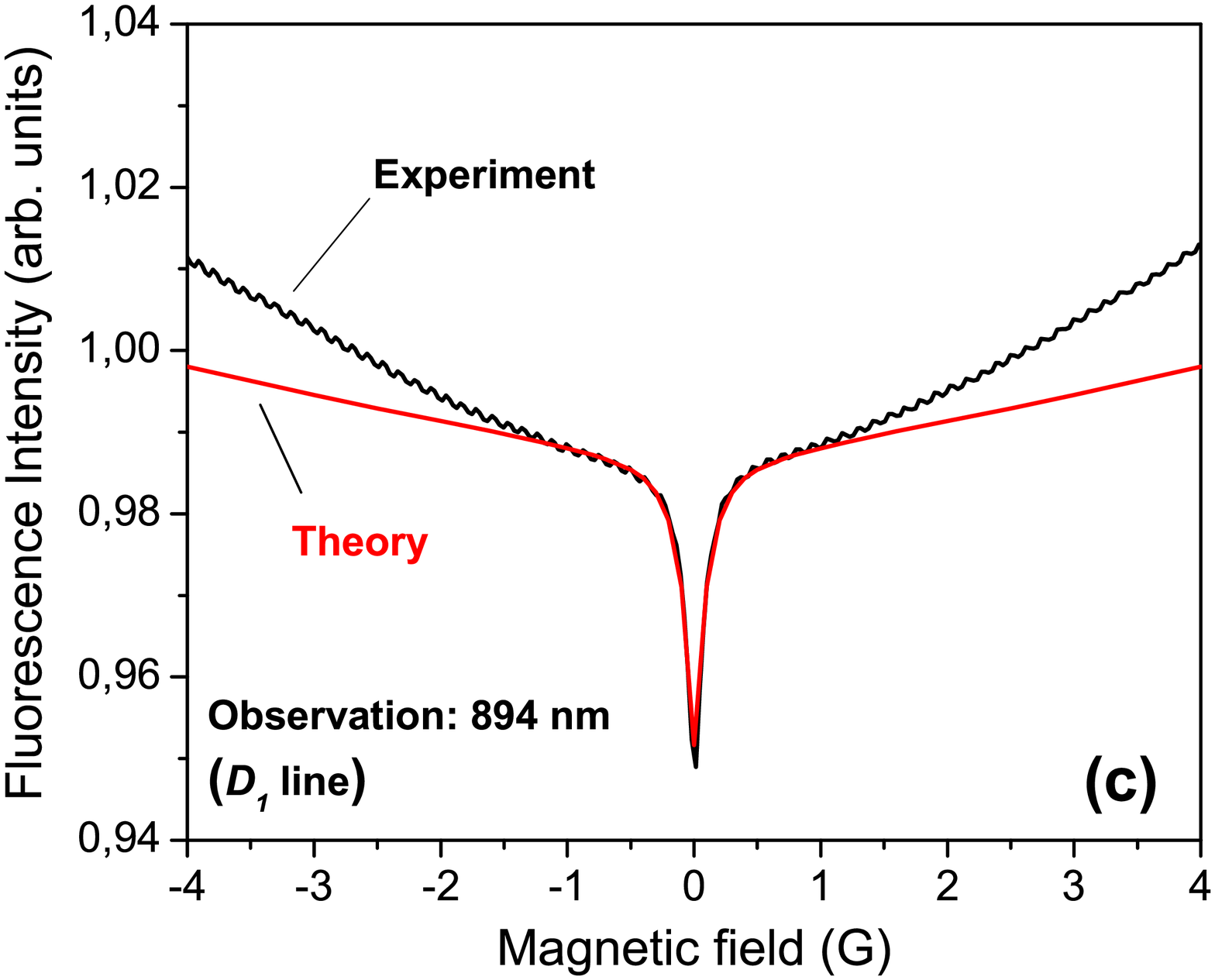}}
	\caption{\label{fig:fg4_np} (Color online) Intensity of the non-polarized fluorescence to the ground state
versus the magnetic field for excitation of the $6S_{1/2}(F_g=4)\rightarrow 7P_{3/2}$ transition at 455 nm. 
(a) Observed fluorescence from the $7P_{3/2}$ (direct), $6P_{3/2}$ (cascade), and $6P_{1/2}$ (cascade) states. 
(b) Theoretical calculations corresponding to the observations in (a). 
(c) Experimental observation and theoretical calculation of the fluorescence from the $6P_{1/2}$ state ($D_1$ transition).
}
\end{figure*}

Figure~\ref{fig:fg4_np} shows similar results as Fig.~\ref{fig:fg3_np}, except that in Fig.~\ref{fig:fg4_np}, the
atoms were excited from the $F_g=4$ ground-state level. One notable difference with the case of excitation from the
$F_g=3$ level is that the resonance shapes did depend on which fluorescence line was observed (see Fig.~\ref{fig:fg4_np}(a).
The theoretical calculations in Fig.~\ref{fig:fg4_np}(b) confirm that the resonance shape, in particular the contrast, 
depends on the fluorescence  line that is observed. 
The theoretical calculations do not reproduce the experimental 
signals extremely well at fields larger than several Gauss.  
However, the theoretical curve in Fig.~\ref{fig:fg4_np}(c) describes 
quite well the narrow portion of the resonance up to a magnetic field of up to about $\pm1$ G. 
For the purpose of modeling the transition, 
we assumed that the laser was tuned to the $F_g=4\rightarrow F_e=4$ transition.

\subsection{Polarized Fluorescence}\label{section:polarized}
\begin{figure*}[htbp]
	\centering
		\resizebox{8cm}{6cm}{\includegraphics{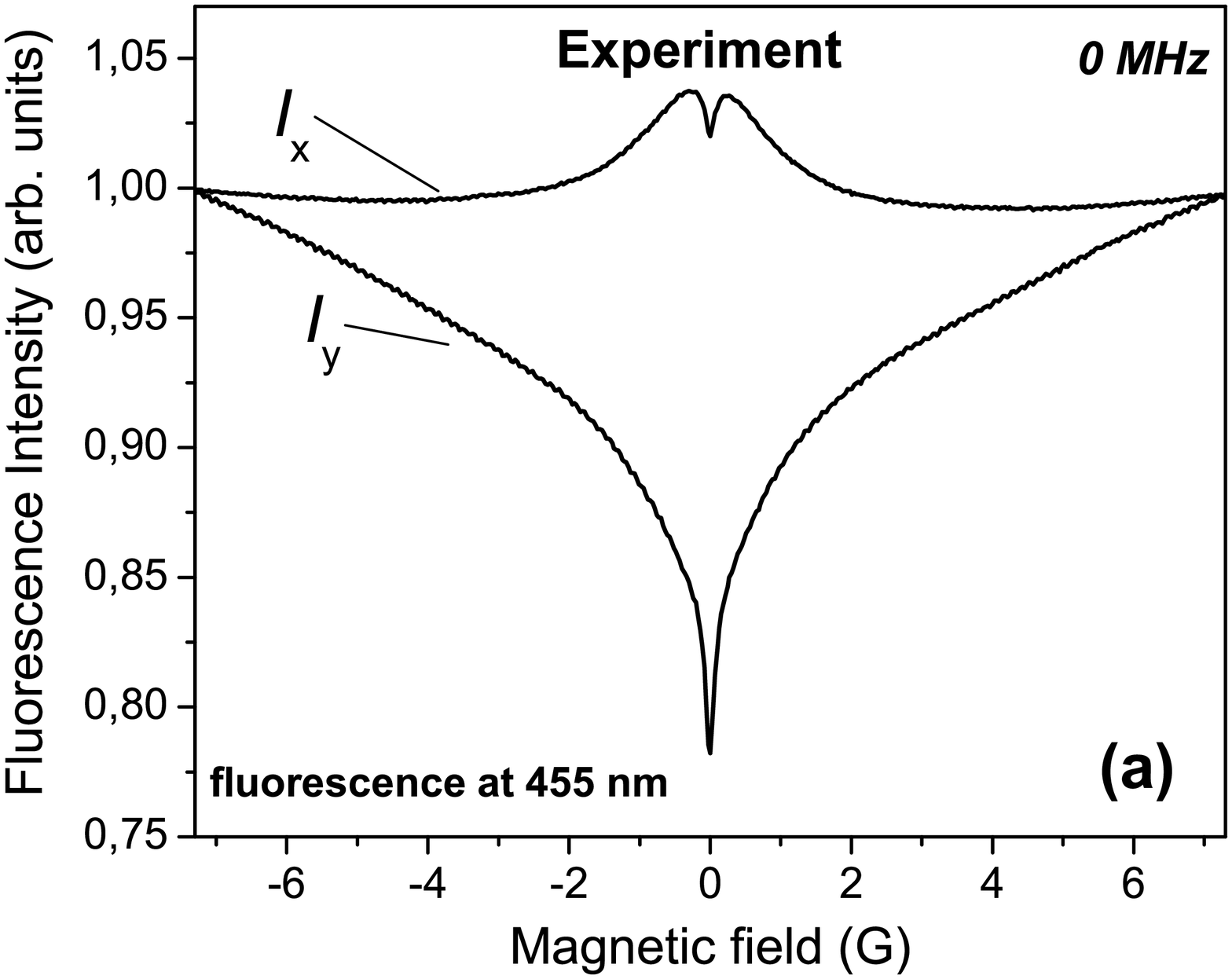}}
		\resizebox{8cm}{6cm}{\includegraphics{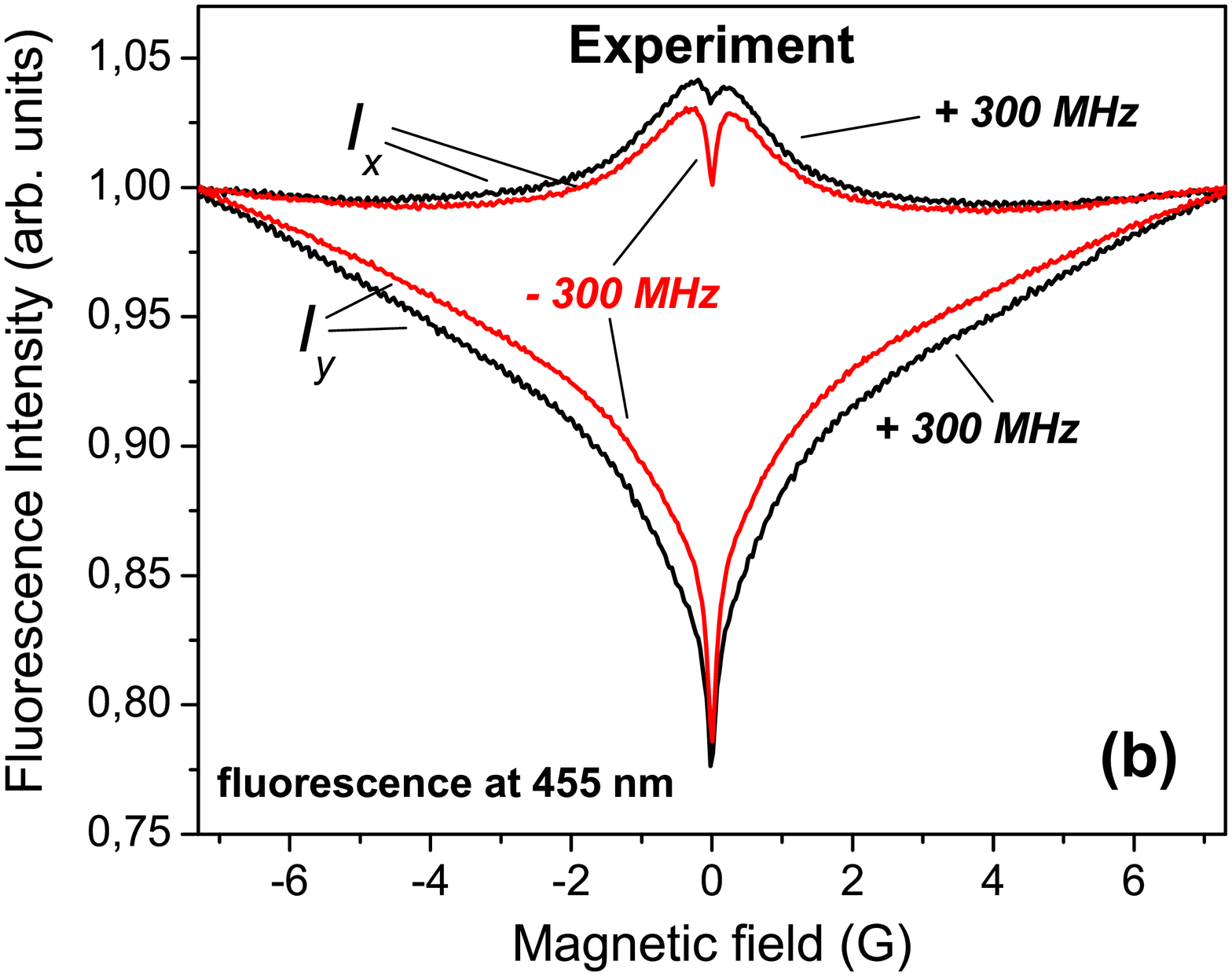}}
		\resizebox{8cm}{6cm}{\includegraphics{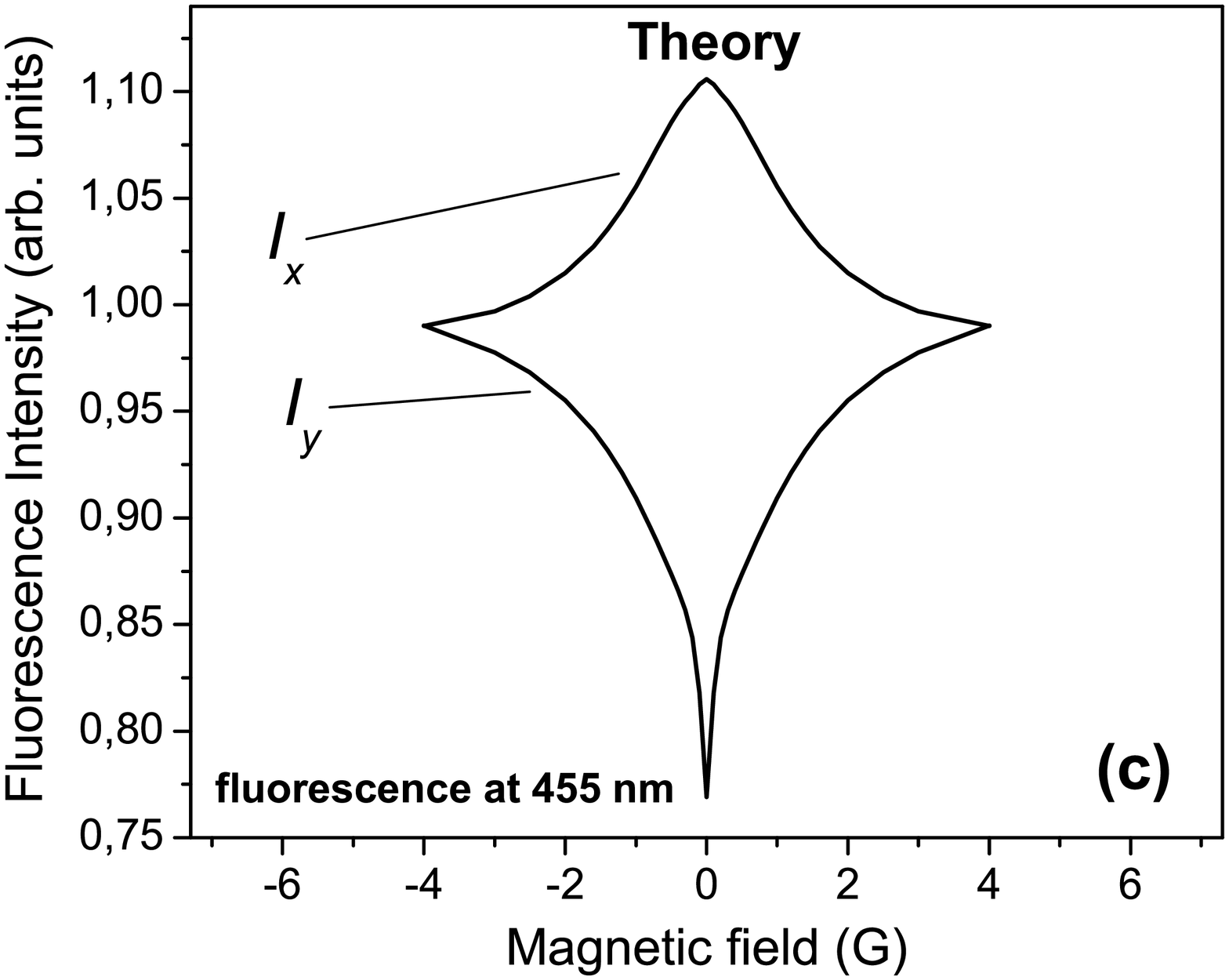}}
	\caption{\label{fig:fg4_455} (Color online) Intensities of the orthogonally polarized components $I_x$ and $I_y$
of the fluorescence from 
the $7P_{3/2}$ state to the ground state versus the magnetic field for excitation from the $F_g=4$ ground-state level for 
various laser detunings from the frequency of maximum observed fluorescence intensity. 
The intensities of the fluorescence with polarization vector parallel or perpendicular to the 
polarization vector of the exciting laser radiation are denoted as $I_{x}$ and $I_{y}$, respectively 
(see Fig.~\ref{fig:geometry}).  
(a) Observations with the exciting laser (455 nm) tuned to the frequency that gave maximum fluorescence (0 detuning). 
(b) Observations with the exciting laser detuned by $\pm 300$ MHz from the frequency at which the observed fluorescence 
intensity was at a maximum. 
(c) Theoretical calculations for the laser tuned exactly to the $F_g=4\rightarrow F_e=4$ transition.
}
\end{figure*}

Figure~\ref{fig:fg4_455} depicts as a function of magnetic field the intensities of two orthogonally polarized 
fluorescence components from the de-excitation of the $7P_{3/2}$ state directly to the ground state for various
values of the laser detuning. The value of zero detuning in Fig.~\ref{fig:fg4_455}(a) was chosen to correspond to 
the frequency at which the value of the observed fluorescence intensity was at a maximum. 
As can be seen in Fig.~\ref{fig:fg4_455}(b), the shape of the measured resonances at magnetic field values up to several
Gauss was not very sensitive to detuning. However, the contrast of the narrow nonlinear magneto-optical resonance in $I_{x}$
near zero magnetic field decreased noticeably as the laser detuning was scanned from -300 
to +300 MHz. The reason is that
at larger detuning the radiation tends to excite more strongly those transitions in which the total 
ground-state angular momentum $F_g$ is less than the total excited state angular momentum $F_e$. Resonances at
transitions with $F_g<F_e$ should be bright rather than dark~\cite{Renzoni:2001,AlnisJPB:2001}. 
Fig.~\ref{fig:fg4_455}(c) shows theoretical calculations for the fluorescence observed directly from the $7P_{3/2}$ state, 
with the  assumption that the laser is tuned exactly to the frequency of the $F_g=4\rightarrow F_e=4$ transition. The dark 
resonance that was observable in $I_x$ in Fig.~\ref{fig:fg4_455}(b) is not apparent in the theoretical calculations. One can 
suppose that the exact frequency of the $F_g=4\rightarrow F_e=4$ transition lies closer to the laser frequency that was 
detuned by +300 MHz from the frequency of maximum fluorescence, than to the laser frequency detuned by -300 MHz, but the 
calculations are too time-consuming to verify this supposition. 

\begin{figure*}[htbp]
	\centering
		\resizebox{8cm}{6cm}{\includegraphics{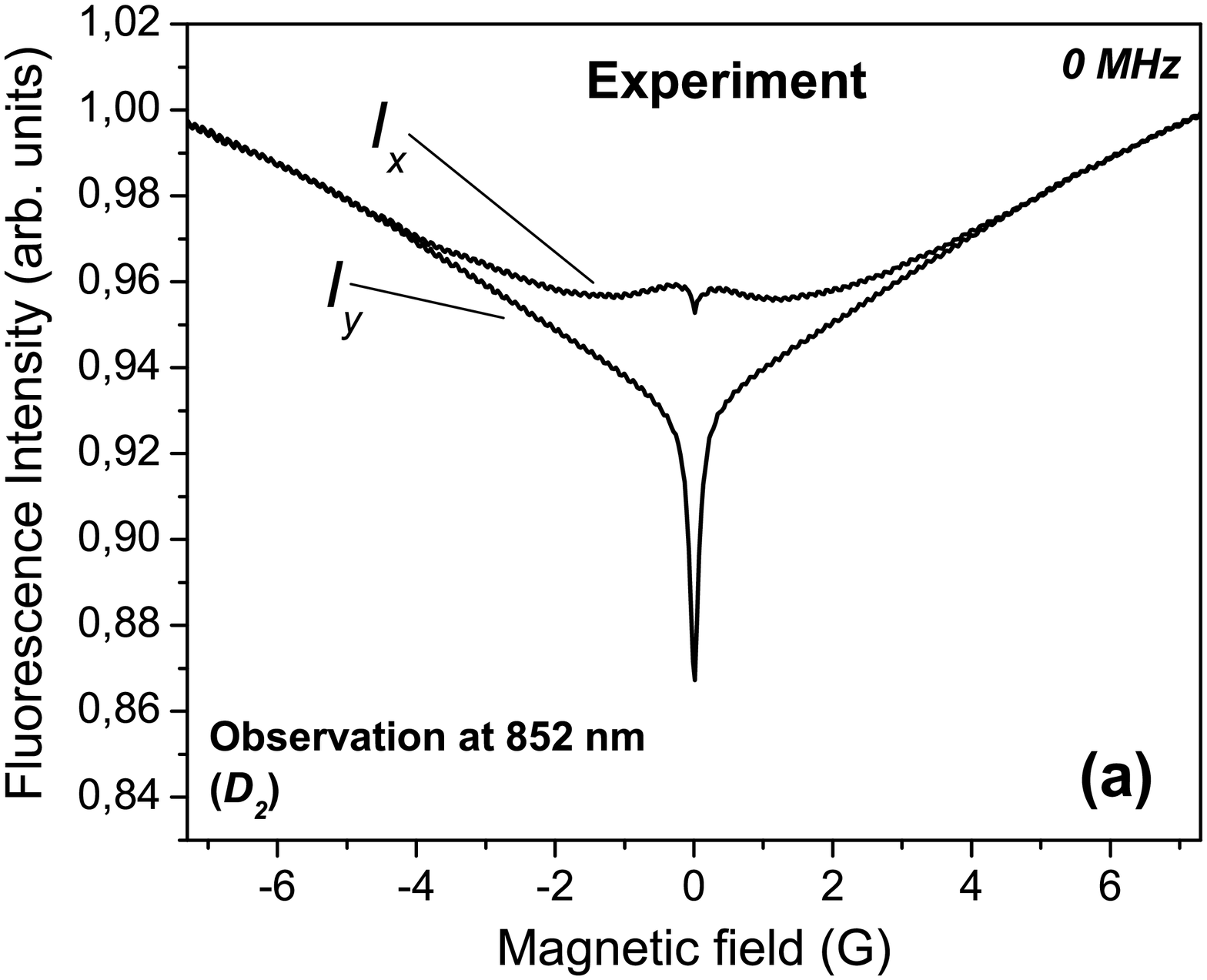}}
		\resizebox{8cm}{6cm}{\includegraphics{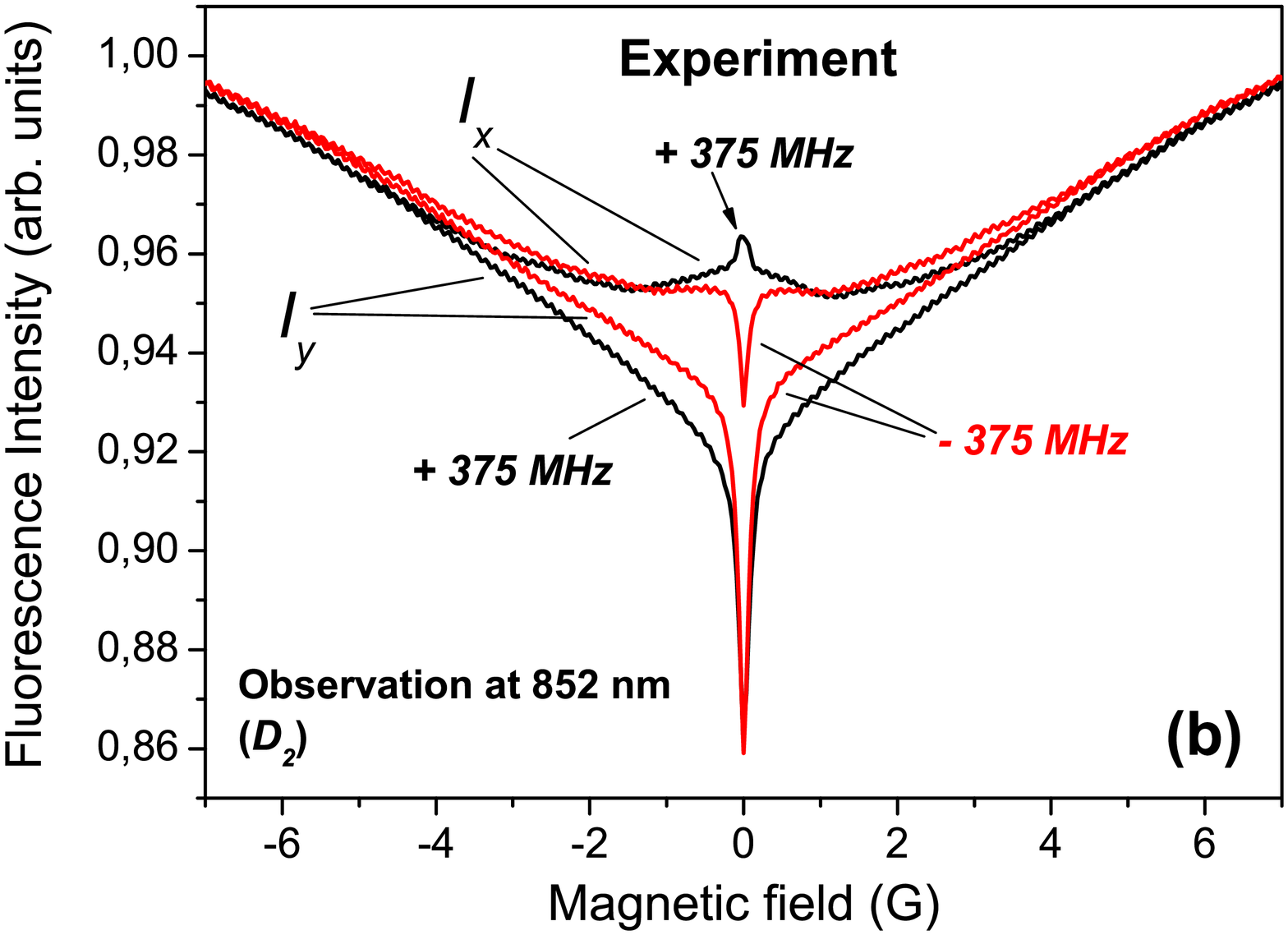}}
		\resizebox{8cm}{6cm}{\includegraphics{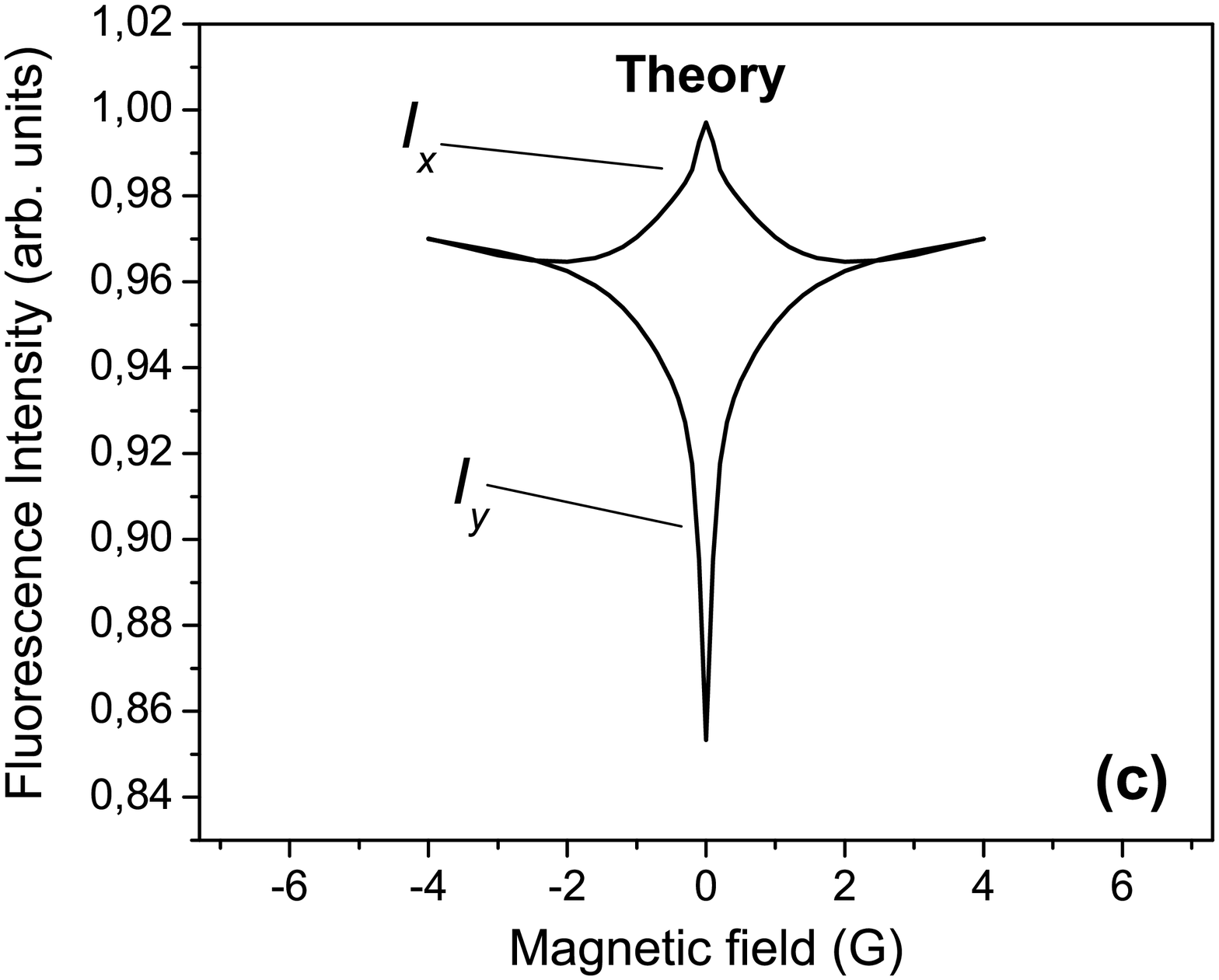}}
	\caption{\label{fig:fg4_d2} (Color online) Intensities of the orthogonally polarized components $I_x$ and $I_y$  
of the fluorescence from 
the $6P_{3/2}$ state to the ground state ($D_2$) versus the magnetic field for excitation from the $F_g=4$ ground-state 
level for various laser detunings from the frequency of maximum observed fluorescence intensity.  
(a) Observations with the exciting laser (455 nm) tuned to frequency that gave maximum fluorescence intensity. 
(b) Observations with the exciting laser detuned by $\pm 375$ MHz from the frequency at which the observed fluorescence 
intensity was at a maximum. 
(c) Theoretical calculations for the laser tuned exactly to the $F_g=4\rightarrow F_e=4$ transition.
}
\end{figure*}

Figure~\ref{fig:fg4_d2} shows the intensity versus magnetic field for two orthogonally polarized 
fluorescence components measured 
from the $6P_{3/2}$ level, which was populated from the $7P_{3/2}$ level via cascades, for various values of the laser
detuning. As in the previous figure, the value of zero detuning [Fig.~\ref{fig:fg4_d2}(a)] was chosen for the laser 
frequency that corresponded to the maximum observed fluorescence intensity. 
Figure~\ref{fig:fg4_d2}(b) shows the results for detunings of -375 and +375 MHz. It is interesting to note that at
a detuning of -375 MHz a narrow dark resonance is observed in $I_{x}$, whereas at a detuning of +375 MHz the resonance 
is bright. As the detuning is increased, the transitions that are excited tend more towards transitions with $F_e>F_g$, 
which is the criterion for a bright resonance. 
Figure~\ref{fig:fg4_d2}(c) 
shows the results of theoretical calculations made with the assumption that the laser is tuned exactly to the 
$F_g=4\rightarrow F_e=4$ transition. The agreement between the curves in Fig.~\ref{fig:fg4_d2}(c) and 
the experiment is not too good, but it is qualitatively correct. In particular, both the 
theoretical curve for zero detuning and the experimental curves for detuning of +375 MHz from the 
frequency of maximum fluorescence show a bright resonance in $I_{x}$.

\subsection{Degree of Polarization of the Fluorescence}\label{section:degree}
\begin{figure*}[htbp]
	\centering
		\resizebox{\columnwidth}{!}{\includegraphics{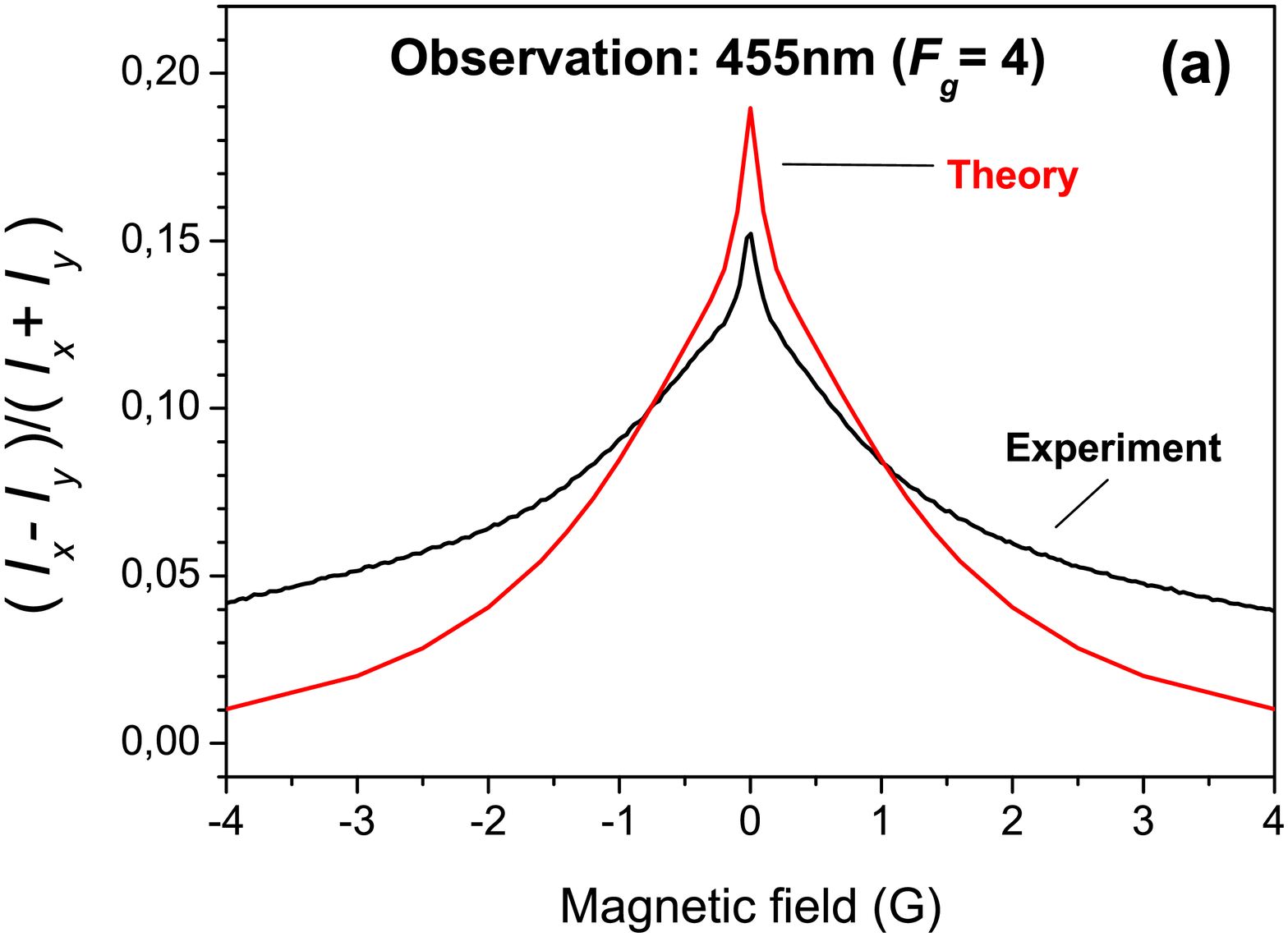}}
		\resizebox{\columnwidth}{!}{\includegraphics{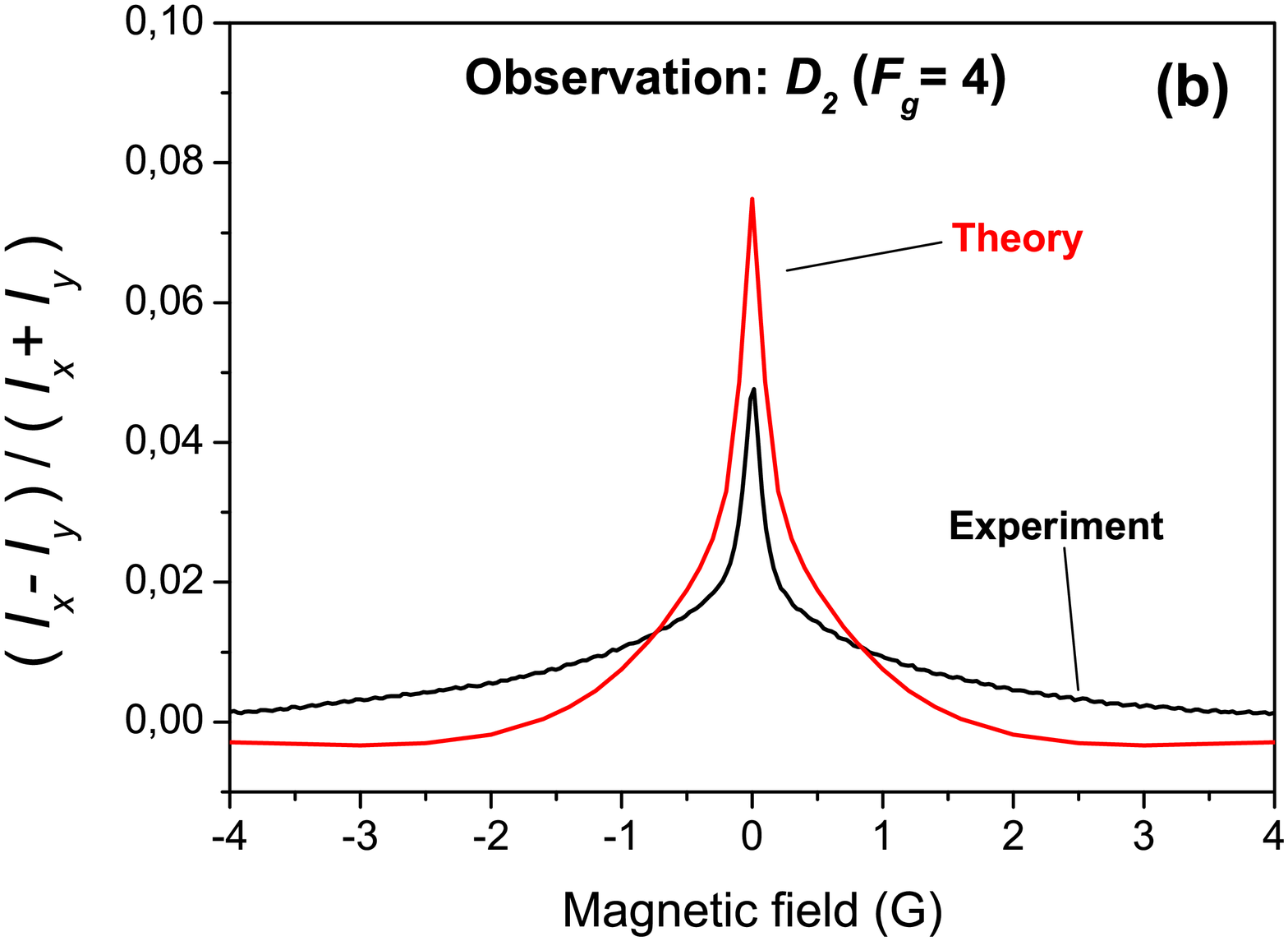}}
	\caption{\label{fig:pol_fg3} (Color online) Polarization degree of the fluorescence [$(I_{x}-I_{y})/(I_{x}+I_{y})$]
for excitation of the $7P_{3/2}$ state from the $F_g=3$ ground-state level as a function of the magnetic field. 
(a) Experiment and theory for observation of the fluorescence to the ground state from the $7P_{3/2}$ state.
(b) Experiment and theory for observation of the fluorescence to the ground state from the $6P_{3/2}$ state 
($D_2$ transition) populated from above via cascade transitions. 
}
\end{figure*}

The degree of polarization of the fluorescence $(I_{x}-I_{y})/(I_{x}+I_{y})$ is plotted as a function 
of the magnetic field when the $7P_{3/2}$ state was excited from the ground-state level with total angular momentum
$F_g=3$ and fluorescence was observed back to the ground state directly
from the $7P_{3/2}$ state [Fig.~\ref{fig:pol_fg3}(a)] or from the $6P_{3/2}$ state [Fig.~\ref{fig:pol_fg3}(b)]. 
The agreement between theory and experiment in this case is quite good over a range of magnetic field values
from $-4$ G to $+4$ G.
However, the theoretical calculation for fluorescence from the $7P_{3/2}$ state suggests that at zero 
magnetic field there should have been a small feature with negative second derivative, 
which was not observed in the experiment.

\begin{figure*}[htbp]
	\centering
		\resizebox{\columnwidth}{!}{\includegraphics{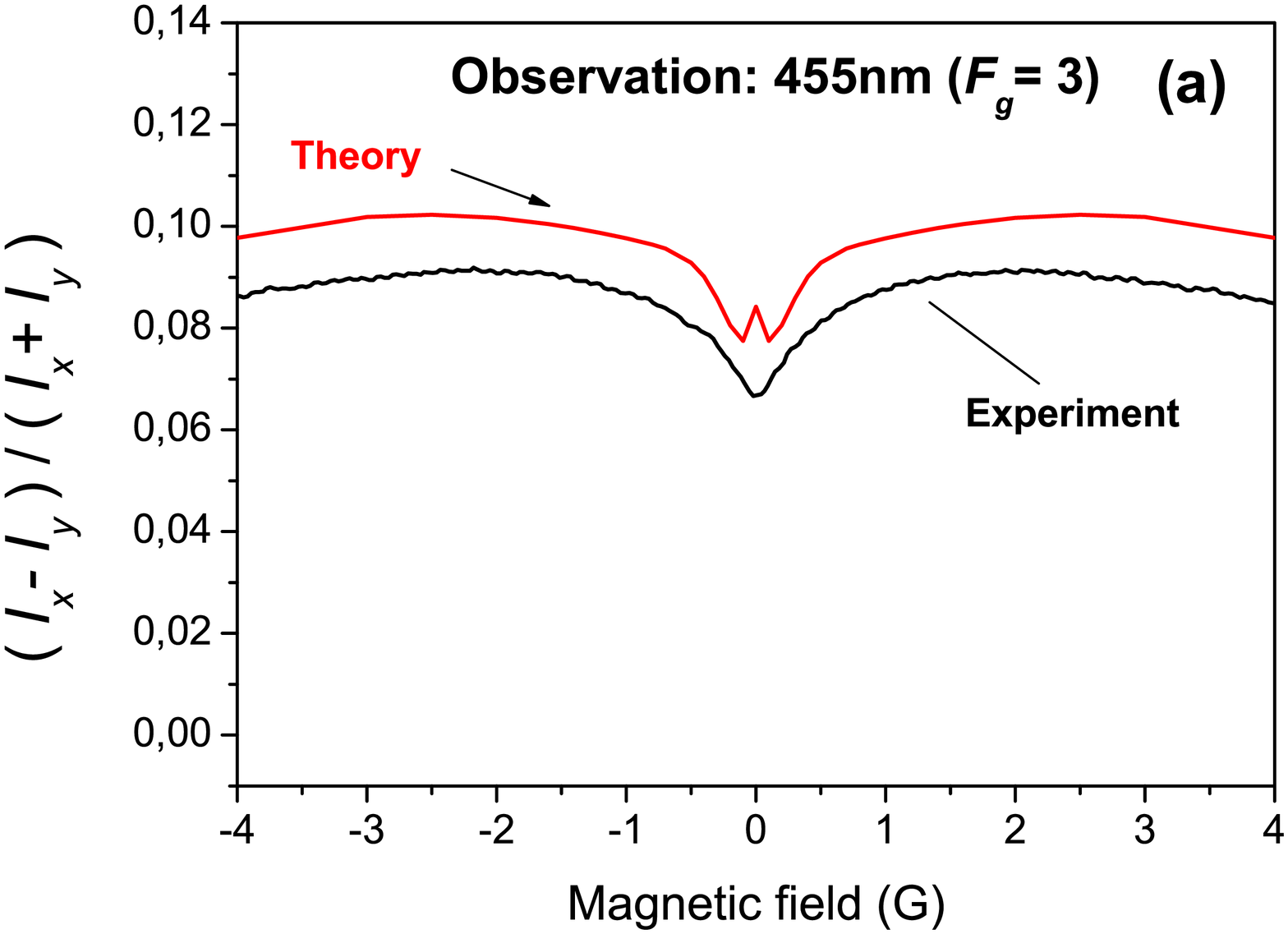}}
		\resizebox{\columnwidth}{!}{\includegraphics{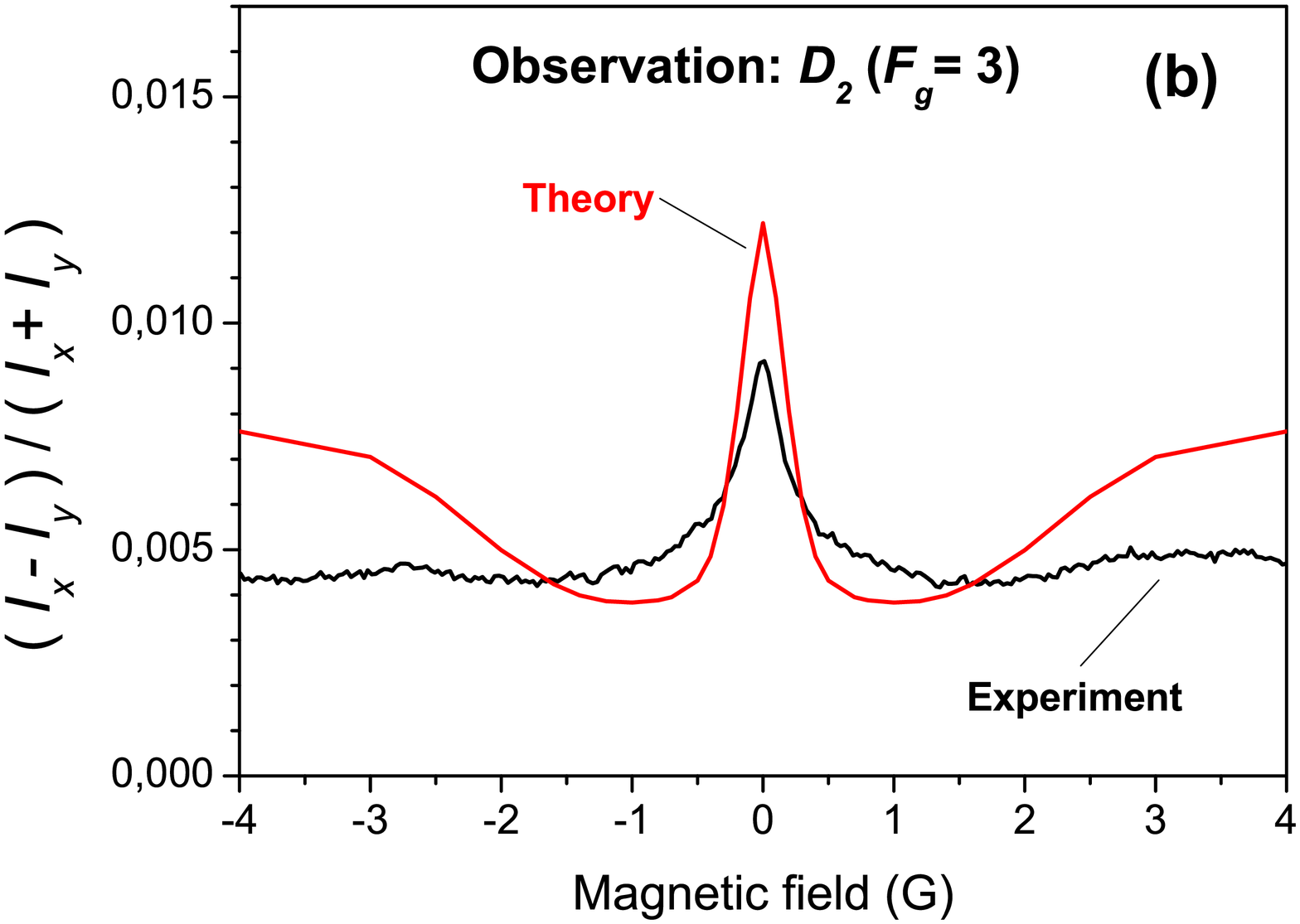}}
	\caption{\label{fig:pol_fg4} (Color online) Polarization degree of the fluorescence [$(I_{x}-I_{y})/(I_{x}+I_{y})$]
from excitation of the $7P_{3/2}$ state from the $F_g=4$ ground-state level as a function of the magnetic field. 
(a) Experiment and theory for observation of the fluorescence to the ground state from the $7P_{3/2}$ state.
(b) Experiment and theory for observation of the fluorescence to the ground state from the $6P_{3/2}$ state 
($D_2$ transition) populated from above via cascade transitions. 
}
\end{figure*}

Figure~\ref{fig:pol_fg4} shows the polarization degree of the fluorescence as a function of magnetic field when  
the $7P_{3/2}$ state was excited from the $F_g=4$ ground-state level. The results for fluorescence 
observed from the $7P_{3/2}$ state to the ground state are shown in Fig.~\ref{fig:pol_fg4}(a), 
while Fig.~\ref{fig:pol_fg4}(b) depicts the results for fluorescence observed from the $6P_{3/2}$ state populated 
from above via cascade transitions. 
Results from experimental observations are compared with the results of theoretical 
calculations. In general, the theoretical calculations qualitatively describe the experimentally measured curves,
although the agreement in the contrast is not very precise. The inevitable depolarization of the experimentally 
measured signals was not taken into account in the theoretical calculations. 

The polarization degree $(I_x-I_y)/(I_x+I_y)$ in Fig.~\ref{fig:pol_fg3} and Fig.~\ref{fig:pol_fg4} is related to the 
polarization moments, which appear in the multipole expansion of the density matrix 
(see, for example, Ref.~\cite{Auzinsh:2010}). 
By comparing the vertical scales of 
Fig.~\ref{fig:pol_fg3}(a) and Fig.~\ref{fig:pol_fg3}(b), one finds that the polarization degree observed from the 
$7P_{3/2}$ level was an order of magnitude higher than the that observed from the $6P_{3/2}$ level when the $7P_{3/2}$ level
was excited from the ground-state sublevel with $F_g=3$ and the $6P_{3/2}$ state was populated by spontaneous 
cascade transitions from the $7P_{3/2}$ state through various intermediate states. It should be noted that the reduction 
in polarization degree depends on the external magnetic field and the shape of the plot of polarization degree versus 
magnetic field markedly differs in Fig.~\ref{fig:pol_fg3}(a) and Fig.~\ref{fig:pol_fg3}(b). 
In the case of excitation from the ground-state sublevel with $F_g=4$, the polarization degree observed from the 
$6P_{3/2}$ state is also smaller than the polarization degree observed from the $7P_{3/2}$ state, 
but only by a factor of three. In this case, the shape of the plots in Fig.~\ref{fig:pol_fg4}(a) and 
Fig.~\ref{fig:pol_fg4}(b) do not differ as dramatically as in the case of excitation from the ground-state 
sublevel with $F_g=3$, although the peak in Fig.~\ref{fig:pol_fg4}(b) is narrower than the peak in 
Fig.~\ref{fig:pol_fg4}(a). 
The reasonable agreement between experimental measurements and theoretical calculations in these figures 
suggests that the density matrix is well known in this system of many levels connected by spontaneous 
cascade transitions.

It should be noted that the transfer of polarization
from one level to another has been studied theoretically in Ref.~\cite{Auzinsh:2009b}. There it was shown that the
maximum polarization moment rank $\kappa$ of an excited state with unresolved hyperfine structure that can be observed through 
fluorescence is $\kappa\le2J_e$. In particular, this means that the polarization degree observed from the $D_1$ line should be 
zero, since a non-zero polarization degree in the fluorescence would imply a polarization rank $\kappa=2$. 
In fact, we measured it to be zero at zero magnetic field 
and less than 0.7\% over the range of magnetic field values from -7G to 7G.

\section{\label{Conclusion:level1}Conclusion}
The ground-state, nonlinear magneto-optical resonances have been observed in the fluorescence from 
the $7P_{3/2}$ state populated by linearly polarized 455 nm laser radiation and from the $6P_{3/2}$ and $6P_{1/2}$ states 
populated via cascade transitions
from the $7P_{3/2}$ state. A theoretical description of these effects has been furnished and compared to 
experimentally measured signals. The theoretical model was based on an earlier model that had been developed 
for $D$-line excitation in alkali metal atoms and was based on the optical Bloch equations with averaging over the Doppler 
profile. This model was modified to take into account the populations of all levels, 
including levels populated by cascade transitions. The model also accounted for the mixing of the magnetic sublevels in an 
external magnetic field, which was significant in the experiment for some of the higher states with small hfs splittings. 
In general, the agreement between the observed signals and the calculated curves was 
surprisingly good, especially taking into account that the experimental parameters were only estimated. In the 
future it would be desirable to take advantage of improved algorithms and more powerful computers to be able to search for 
the values of the experimental parameters that could not be measured explicitly by varying the parameter values in the 
model. Cascade techniques are interesting 
because they provide a way to observe magneto-optical resonances in fluorescence at a frequency far removed from the 
exciting laser radiation.

\begin{acknowledgments}
We thank Stefka Cartaleva for drawing our attention to this interesting research question and Robert Kalendarev 
for preparing the cesium cell. 
This project was carried out with  support from  the Latvian Science Council Grant 
No. LZP 09.1567 and the State Research Program Grant No. 2010/10-4/VPP-2/1 ``Development of innovative multifunctional 
materials, signal processing and information technology for competitive, science intensive products'', 
and the ESF project Nr. 2009/0223/1DP/1.1.1.2.0./09/APIA/VIAA/008. 
L.~K. acknowledges support from the ESF project Nr. 2009/0138/1DP/1.1.2.1.2./09/IPIA/VIAA/004.
\end{acknowledgments}





\bibliography{cascades}







\end{document}